\documentclass[twocolumn,floatfix,aps,prd,showpacs]{revtex4}

\usepackage{graphicx}
\usepackage{bm}
\usepackage{epsf}

\newcommand{\beq}{\begin{equation}}
\newcommand{\eeq}{\end{equation}}
\newcommand{\beqn}{\begin{eqnarray}}
\newcommand{\eeqn}{\end{eqnarray}}
\newcommand{\pa}{\partial}

\newcommand{\varep}{\varepsilon}

\newcommand{\cC}{{\cal{C}}}
\newcommand{\brr}{{\mbox{\boldmath$r$}}}

\begin{document}

\title{Gravitational waves from black hole-neutron star binaries I: 
Classification of waveforms}

\author{$^1$Masaru Shibata, $^2$Koutarou Kyutoku, $^3$Tetsuro Yamamoto,
$^4$Keisuke Taniguchi}

\affiliation{$^1$Yukawa Institute of Theoretical Physics, 
Kyoto University, Kyoto, 606-8502, Japan\\
$^2$Department of Physics, University of Tokyo, Tokyo 113-0033, Japan \\
$^3$Yugen Club, Toyama, Shinjuku, Tokyo 162-0052, Japan \\
$^4$Department of Physics, University of Wisconsin-Milwaukee,
P.O. Box 413, Milwaukee, Wisconsin 53201
}


\begin{abstract}
Using our new numerical-relativity code {\tt SACRA}, long-term
simulations for inspiral and merger of black hole (BH)-neutron star
(NS) binaries are performed, focusing particularly on gravitational
waveforms. As the initial conditions, BH-NS binaries in a
quasiequilibrium state are prepared in a modified version of the
moving-puncture approach. The BH is modeled by a nonspinning moving
puncture and for the NS, a polytropic equation of state with
$\Gamma=2$ and the irrotational velocity field are employed.  The mass
ratio of the BH to the NS, $Q=M_{\rm BH}/M_{\rm NS}$, is chosen in the
range between 1.5 and 5.  The compactness of the NS, defined by ${\cal
C}=GM_{\rm NS}/c^2R_{\rm NS}$, is chosen to be between 0.145 and
0.178.  For a large value of $Q$ for which the NS is not tidally
disrupted and is simply swallowed by the BH, gravitational waves are
characterized by inspiral, merger, and ringdown waveforms. In this
case, the waveforms are qualitatively the same as that from BH-BH
binaries.  For a sufficiently small value of $Q \alt 2$, the NS may be
tidally disrupted before it is swallowed by the BH. In this case,
the amplitude of the merger and ringdown waveforms is very low, and
thus, gravitational waves are characterized by the inspiral waveform and
subsequent quick damping. The difference in the merger and ringdown
waveforms is clearly reflected in the spectrum shape and in the
``cut-off'' frequency above which the spectrum amplitude steeply
decreases.  When an NS is not tidally disrupted (e.g., for $Q=5$),
kick velocity, induced by asymmetric gravitational wave emission, 
agrees approximately with that derived for the merger of BH-BH
binaries, whereas for the case that the tidal disruption occurs, the
kick velocity is significantly suppressed.
\end{abstract}
\pacs{04.25.D-, 04.30.-w, 04.40.Dg}

\maketitle

\section{Introduction}

The coalescence of black hole (BH)-neutron star (NS) binaries is one
of the most promising sources for kilometer-size laserinterferometric
gravitational-wave detectors such as LIGO \cite{LIGO} and VIRGO
\cite{VIRGO}.  A statistical study based on the stellar evolution
synthesis suggests that the detection rate of gravitational waves from
BH-NS binaries will be about 1/20--1/3 of the rate expected for the merger
of the NS-NS binaries in the universe \cite{BHNS,Bel}. Then, 
the detection rate of such systems will be $\sim 0.5$--50 
events per year for advanced detectors such as advanced-LIGO 
\cite{advLIGO}, and hence, the detection is expected to be achieved in
the near future. For clarifying the nature of the sources of
gravitational waves and for extracting their physical information, 
theoretical templates of gravitational waves are necessary
for the data analysis. For theoretically computing gravitational
waveforms for the late inspiral and merger phases of BH-NS binaries,
numerical relativity is the unique approach.

The final fate of coalescing BH-NS binaries is divided into two cases
depending primarily on the BH mass. When the BH mass is small enough,
the companion NS will be tidally disrupted before it is swallowed by
the BH. By contrast, when the BH mass is large enough, the NS will be
swallowed by the BH without tidal disruption. The latest study for the
BH-NS binary in a quasiequilibrium indicates that the tidal disruption
of the NSs by a nonspinning BH will occur for the case that the BH mass
is $\alt 4M_{\odot}$, for the hypothetical mass and radius of the NS,
$M_{\rm NS}\sim 1.35M_{\odot}$ and $R_{\rm NS}\sim 11$--12 km,
respectively \cite{TBFS,TBFS2,FKPT}.  The tidally disrupted NSs may
form a disk or a torus around the BH if the tidal disruption occurs
outside the innermost stable circular orbit (ISCO). A system
consisting of a rotating BH surrounded by a massive, hot torus has
been proposed as one of the likely sources for the central engine of
gamma-ray bursts with a short duration \cite{grb2} (hereafter
SGRBs). Hence, the merger of a low-mass BH and its companion NS can be
a candidate of the central engine. According to the observational
results by the {\it Swift} and {\it HETE-2} satellites \cite{Swift},
the total energy of the SGRBs is larger than $\sim 10^{48}$ ergs, and
typically $10^{49}$--$10^{50}$ ergs. The studies of hypercritical,
neutrino-dominated accretion disks around a BH suggest that disk mass
should be larger than $\sim 0.01M_{\odot}$ for providing such high
energy for generating gamma-ray via neutrino process
\cite{GRBdisk,GRBdisk1}.  The question is whether or not the mass and
thermal energy of the disk (torus) are large enough for driving the
SGRBs of the huge total energy. Numerical-relativity simulation plays
an important role for answering this question.

In the last three years, general relativistic numerical simulations
for the inspiral and merger of BH-NS binaries have been performed by
three groups \cite{SU06,ST08,ILLINOIS,SACRA,CORNELL}. However, most of
these previous works should be regarded as preliminary ones because
the simulations were performed only for a short time scale: In the
early works \cite{SU06,ST08,ILLINOIS}, the inspiral motion is followed
for at most two orbits. With such short-term simulations, the orbit is
not guaranteed to be quasicircular at the onset of the merger, and
hence, the obtained results are not likely to be very realistic
because of the presence of the unrealistic eccentricity and incorrect
approaching velocity.  Moreover, by such incorrect conditions at the
onset of the merger, the final outcome such as mass and spin of the BH
and the physical condition for the disk surrounding the BH is not likely
to be computed correctly, although a rough qualitative feature of the
merger mechanism and gravitational waveforms have been found from
these works. In the latest works \cite{SACRA,CORNELL}, the inspiral
motion is followed for $\sim 4$ orbits, but in each of these works,
the simulation was performed only for one model. More systematic study
is obviously necessary to clarify the quantitative details about the
merger process, the final outcome, and gravitational waveforms. 

In this paper, we report our latest work in which a long-term
simulation of BH-NS binaries is performed for a wide variety of BH
masses, NS masses, and NS radii for the first time. In the present
simulation, the inspiral motion is followed for 4--7 orbits. With this
setting, the eccentricity of the last inspiral motion appears to be
negligible and the approaching velocity at the onset of the merger is
correctly taken into account. Furthermore, we systematically choose
the BH and NS masses and NS radii for a wide range, and as a result,
it becomes possible to clarify the dependence of the merger process,
final outcome, and gravitational waveforms on the mass ratio of the
binary and the compactness of the NS. One drawback in the present work
is that the NSs are modeled by a simple equation of state (EOS).
However, the various features found in this paper will qualitatively
hold irrespective of the EOS and thus the present work will be an
important first step toward more detailed simulation in the near
future in which the NSs are modeled by more realistic EOSs.

The paper is organized as follows.  Section II summarizes the initial
conditions chosen in this paper.  Section III briefly describes the
formulation and methods for the numerical simulation. Section IV
presents the numerical results of the simulation focusing primarily on
the dependence of the merger mechanism and gravitational waveforms on
the mass ratio and radius of NSs.  Section V is devoted to a summary.
Throughout this paper, the geometrical units of $c=G=1$, where $c$ and
$G$ are the speed of light and gravitational constant, are used,
otherwise stated.  The irreducible mass of a BH, rest mass of an NS,
gravitational mass and circumferential radius of an NS in isolation,
Arnowitt-Deser-Misner (ADM) mass of system, and sum of the BH and NS
mass (often referred to as the total mass) are denoted by $M_{\rm
BH}$, $M_*$, $M_{\rm NS}$, $R_{\rm NS}$, $M$, and $m_0 (=M_{\rm
BH}+M_{\rm NS})$, respectively. The mass ratio of the binary and the
compactness of the NS are defined by $Q \equiv M_{\rm BH}/M_{\rm NS}$
and ${\cal C}\equiv GM_{\rm NS}/c^2R_{\rm NS}$, respectively.  Note
that $M_{\rm BH}$ is equal to the ADM mass of the BH in isolation for
a nonspinning BH. Greek indices and Latin indices denote the spacetime
and space components, respectively; Cartesian coordinates are used 
for the spatial coordinates. 

\section{Initial condition}

BH-NS binaries in a quasiequilibrium state are employed as an initial
condition of the numerical simulation. Following our previous papers
\cite{SU06,ST08}, the quasiequilibrium state is computed in the
moving-puncture framework \cite{BB,BB2}. The formulation in this
framework is slightly different from that in the excision framework
which is adopted in most of the previous
works~\cite{GRAN,TBFS0,TBFS,TBFS2,FKPT}, although both formulations
are based on the conformal-flatness formalism for the three-metric
\cite{IWM}. In the present work, we adopt the same basic equations as
those in Ref.~\cite{SU06}. However, we change the condition for
defining the center of mass of system to improve the quality of the
quasiequilibrium (specifically, to reduce the orbital eccentricity).
To clarify which part is changed, we summarize the basic equations and
method for determining the quasiequilibrium state again in the
following.  Detailed numerical solutions and their properties are
presented in an accompanied paper \cite{KST}, to which the reader may
refer for details.

\subsection{Formulation}

If the orbital separation of a BH-NS binary is large enough, the time
scale of gravitational-wave emission ($\tau_{\rm GW}$) is much longer
than the orbital period ($P_0$).  In the present work, we follow the
inspiral motion of the BH-NS binaries for more than 4 orbits
(typically $\agt 5$ orbits). This implies that the initial binary
separation is always large enough to satisfy the condition, $\tau_{\rm
GW} \gg P_0$. Thus, the binaries initially given should be in a
quasicircular orbit (i.e., the BH and NS are approximately in an
equilibrium in the comoving frame with an angular velocity
$\Omega$). To obtain such a state, we may assume the presence of a
helical Killing vector around the center of mass of the system as
follows, \beq \ell^{\mu}=(\pa_t)^{\mu} +\Omega (\pa_{\varphi})^{\mu},
\eeq where $\Omega$ denotes the orbital angular velocity.

We assume that the NSs have the irrotational velocity field because it
is believed to be realistic for the BH-NS binaries in close orbits
\cite{KBC}. For the fluid of the irrotational velocity field, the
Euler and continuity equations reduce to a first integral of motion
and an elliptic-type equation for the velocity potential in the
presence of the helical Killing vector \cite{ST}. As a result, the
density profile and velocity field are determined by solving these
hydrostatic equations.

For computing a solution of the geometric variables of a
quasiequilibrium state, we employ the so-called conformal-flatness
formalism for the three-geometry \cite{IWM}. In this formalism, a
solution is obtained by solving Hamiltonian and momentum constraint
equations, and an equation for the lapse function ($\alpha$) which is
derived by imposing the maximal slicing condition as $K=0=\pa_t K$
where $K$ is the trace part of the extrinsic curvature $K_{ij}$.
These equations lead to the equations for the conformal factor $\psi$, a
rescaled tracefree extrinsic curvature $\hat A_{i}^{~j} \equiv \psi^6
K_{i}^{~j}$, and a weighted lapse $\Phi\equiv \alpha\psi$ as
\beqn 
&&\Delta \psi = -2\pi \rho_{\rm H} \psi^5 -{1 \over 8} \hat A_{i}^{~j}
\hat A_{j}^{~i}\psi^{-7}, \label{ham2} \\ && \hat A^{~j}_{i~,j} = 8\pi
J_i \psi^6,\label{mom2} \\ && \Delta \Phi = 2\pi \Phi \Big[\psi^4
(\rho_{\rm H} + 2 S) +{7 \over 16\pi} \psi^{-8}\hat A_{i}^{~j} \hat
A_{j}^{~i}\Big],
\label{alpsi}
\eeqn
where $\Delta$ denotes the flat Laplacian, 
$\rho_{\rm H}=\rho h (\alpha u^t)^2-P$, $J_i=\rho h \alpha u^t u_i$, 
and $S=\rho h [(\alpha u^t)^2-1]+3P$. $\rho$ is the rest-mass density, 
$h$ is the specific enthalpy defined by $1+\varepsilon+P/\rho$, 
$\varep$ is the specific internal energy, $P$ is the pressure, and 
$u^{\mu}$ is the four-velocity. 

For the relation among $\rho$, $\varep$, and $P$, we adopt the
polytropic EOS as,
\beq
P=\kappa \rho^{\Gamma},\label{EOSEOS}
\eeq 
where $\kappa$ is the adiabatic constant and $\Gamma$ the adiabatic 
index for which we choose 2 in this paper. In this EOS, 
$\varep$ is given by $P/(\Gamma-1)\rho$. 

We solve the elliptic equations (\ref{ham2})--(\ref{alpsi}) 
in the moving-puncture framework \cite{BB,BB2,BB4}. Assuming that 
the puncture is located at $\brr_{\rm P}(=x_{\rm P}^k)$, 
we set $\psi$ and $\Phi$ as 
\beqn
\psi=1+{M_{\rm P} \over 2 r_{\rm BH}} + \phi
\ \ \mbox{and} \ \  
\Phi=1 - {M_{\Phi} \over r_{\rm BH}} + \eta, \label{eq6}
\eeqn
where $M_{\rm P}$ and $M_{\Phi}$ are positive constants of mass dimension,
and $r_{\rm BH}=|x^k_{\rm BH}|$ ($x^k_{\rm BH}=x^k-x^k_{\rm P}$).  Then, 
substituting Eq. (\ref{eq6}) into Eqs. (\ref{ham2}) and (\ref{alpsi}), 
elliptic equations for new non-singular functions $\phi$ and $\eta$ are
derived. 

The mass parameter $M_{\rm P}$ may be arbitrarily given, and thus, 
it is appropriately chosen to obtain a desired BH mass. For a 
given value of $M_{\rm P}$, $M_{\Phi}$ is determined by the virial 
relation, which should hold for stationary spacetimes 
(e.g., Ref.~\cite{vir}), as 
\beqn
\oint_{r \rightarrow \infty} \pa_i \Phi dS^i=
-\oint_{r \rightarrow \infty} \pa_i \psi dS^i=2\pi M_0, 
\eeqn
where $M_0$ is the initial ADM mass of the system. 

For a solution obtained in this formalism, $\alpha$ always becomes
negative near the puncture (approximately, inside apparent horizon),
but such lapse is not favorable for numerical simulation.  Following
previous papers \cite{SU06,ST08}, thus, the initial condition for
$\alpha$ is appropriately modified so as to satisfy the condition of
$\alpha > 0$ everywhere.

Equation (\ref{mom2}) is rewritten by setting 
\beqn
\hat A_{ij}(=\hat A_i^{~k}\delta_{jk})
=W_{i,j}+W_{j,i}-{2 \over 3}\delta_{ij} \delta^{kl}
W_{k,l}+K^{\rm P}_{ij},\label{hataij}
\eeqn
where $W_i(=W^i)$ denotes an auxilary three-dimensional function and 
$K^{\rm P}_{ij}$ denotes a weighted extrinsic curvature 
associated with the linear momentum of a BH; 
\beqn
K^{\rm P}_{ij}={3 \over 2 r_{\rm BH}^2}\biggl(n_i P_j +n_j P_i
+(n_i n_j-\delta_{ij}) P_k n^k \biggr). 
\eeqn
Here, $n_k=n^k=x^k_{\rm BH}/r_{\rm BH}$ and $P_i(=P^i)$ 
denotes the linear momentum of the BH, determined from the condition 
that the total linear momentum of system is zero as 
\beqn
P_i=-\int J_i \psi^6 d^3x. \label{P}
\eeqn
The right-hand side of Eq. (\ref{P}) denotes the linear momentum 
of the companion NS. 
Then, the total angular momentum of the system is derived from
\beqn
J=\int J_{\varphi} \psi^6 d^3x + \epsilon_{zjk} x_{\rm P}^j P^k,
\eeqn
where we assume that the $z$ axis is the axis of orbital rotation. 

Substituting Eq. (\ref{hataij}) into Eq. (\ref{mom2}),
an elliptic equation for $W_i$ is derived in the form 
\beqn
\Delta W_i + {1 \over 3}\pa_i \pa_k W^k=8\pi J_i \psi^6. \label{weq}
\eeqn
Denoting $W_i=7 B_i - (\chi_{,i}+B_{k,i} x^k)$ where $\chi$ and $B_i$
are auxiliary functions,
we decompose Eq. (\ref{weq}) into two linear elliptic equations 
\beq
\Delta B_i = \pi J_i \psi^6~~{\rm and}~~\Delta \chi= -\pi J_i x^i \psi^6.
\eeq

To determine a BH-NS binary in quasiequilibrium states, in addition, 
we have to determine the shift vector, $\beta^i$. 
The reason for this is that $\beta^i$ appears in the hydrostatic equations 
\cite{KST}. In the conformal-flatness formalism, the relation between 
$\beta^i$ and $\hat A_{ij}$ is written as 
\beqn
\delta_{jk} \pa_i \beta^k +\delta_{ik} \pa_j \beta^k
-{2\over 3}\delta_{ij} \pa_k \beta^k ={2\alpha \over \psi^6} \hat A_{ij}.
\label{eq13}
\eeqn
Operating $\delta^{jl}\pa_l$ to this equation, an elliptic equation 
is derived 
\beqn
\Delta \beta^i + {1 \over 3} \delta^{ik} \pa_k\pa_j\beta^j
=2 \pa_j (\alpha \psi^{-6}) \hat A^{ij}+16\pi\alpha J_j \delta^{ij}.
\label{betaeq}
\eeqn
For $\hat A_{ij}$ in the right-hand side of this equation, we
substitute the relation of Eq. (\ref{hataij}) [not Eq. (\ref{eq13})].
As a result, no singular term appears in the right-hand side of
Eq. (\ref{betaeq}), and thus, $\beta^i$ is solved in the same manner as
that for $W_i$.

In this formulation, the elliptic equations for the
gravitational-field components, $\phi$, $\eta$, $B_i$, $\chi$, and
$\beta^i$, and the velocity potential have to be solved.  For a
numerical solution of them, we use the LORENE library \cite{LORENE},
by which a high-precision numerical solution can be computed using the
spectral method. We note that in the puncture framework, we do not
basically have to impose an inner boundary condition around the BH. In
this point, the moving-puncture framework differs from the excision
framework, and this may be a demerit of the moving-puncture framework
because a physical condition may not be imposed for the BH. However,
this may be also a merit of this framework because there remains a
degree of freedom which can be used to adjust the property of the
quasiequilibrium to a desired state, as mentioned in the following.

\begin{figure*}[t]
\epsfxsize=3.2in
\leavevmode
(a)\epsffile{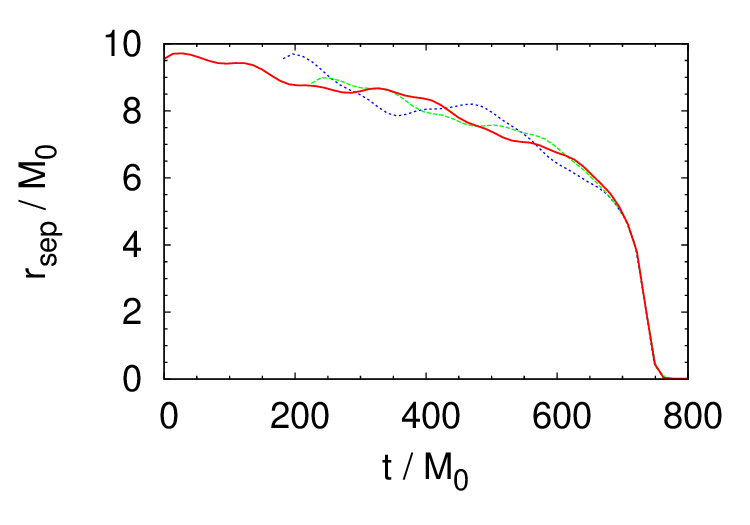}
\epsfxsize=3.2in
\leavevmode
~~(b)\epsffile{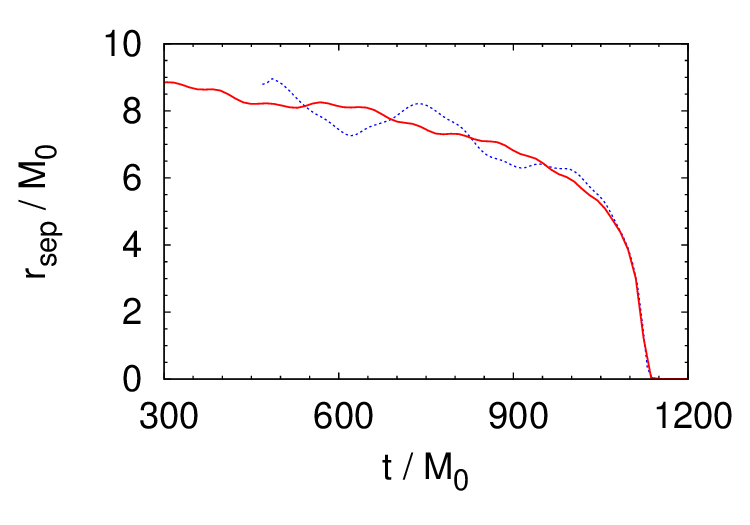}
\vspace{-2mm}
\caption{Evolution of orbital separation (a) for model M20.145 and (b)
  for model M50.145. The solid and dotted curves denote the results
  with the initial conditions obtained in the 3PN-J and
  $\beta^{\varphi}$ conditions, respectively. The dashed curve in
  panel (a) denote the result for model M20.145N. To align the curves
  at the onset of the merger, the time is appropriately shifted for the
  results of M20.145N, M20.145b, and M50.145b.  Note that the merger
  sets in when the orbital separation becomes $\sim 5M_0$.  The
  binaries for models M20.145 and M50.145 spend in the inspiral phase
  for $\sim 5$ and $7.5$ orbits, respectively (see Sec. IV), whereas
  those for models M20.145b and M50.145b spend only for $\sim 4$ and 
  5.25 orbits, respectively.
\label{FIG1}}
\end{figure*}

The final remaining task in the moving-puncture framework is to
determine the center of mass of system. The issue in this framework is
that we do not have any natural physical condition for determining it.
(By contrast, the condition is automatically derived in the excision
framework \cite{TBFS2}, although it is not clear whether the condition
is really physical and whether the resulting quasiequilibrium is a
quasicircular state; see, e.g., Ref.~\cite{BERTI} which assesses the
circularity of the quasiequilibrium.)  In our first paper \cite{SU06},
we employed a condition that the dipole part of $\psi$ at spatial
infinity is zero. However, in this method, the angular momentum
derived for a close orbit of $m_0 \Omega \agt 0.03$ is by $\sim 2\%$
smaller than that derived by the third post-Newtonian (3PN)
approximation \cite{Luc}. (Note $m_0=M_{\rm BH} + M_{\rm NS}$.)
Because the 3PN approximation should be an excellent approximation of
general relativity for describing a binary of a fairly distant orbit
as $m_0 \Omega \approx 0.03$, we should consider that the obtained
initial data deviates from the true quasicircular state, and thus, the
initial orbit would be eccentric.  Such initial condition is not
suitable for quantitatively accurate numerical simulation of
the inspiraling BH-NS binaries.

In the subsequent work \cite{ST08}, we adopted a condition that the
azimuthal component of the shift vector $\beta^{\varphi}$ at the
location of the puncture ($\brr=\brr_{\rm P}$) is equal to $-\Omega$;
i.e., we imposed a corotating gauge condition at the location of the
puncture. In the following, we refer to this condition as
``$\beta^{\varphi}$ condition''.  This is slightly better than the
original condition, but the angular momentum derived for a close orbit
of $m_0 \Omega \agt 0.03$ is still by $\agt 1\%$ smaller than that
derived by the 3PN relation for a large mass ratio $Q \geq 2$ (see
Ref.~\cite{KST} for detailed numerical results). The disagreement is
larger for the larger mass ratio. As a result of this, the initial
condition is likely to deviate from the true quasicircular state and
hence the initial orbital eccentricity is not also negligible (see
Sec. IV C of Ref.~\cite{SACRA} for numerical evolution of such initial
data), in particular, for binaries of large mass ratios. This also
suggests that the $\beta^{\varphi}$ condition is not suitable for
deriving a realistic quasicircular state.

If the simulations are started with a sufficiently large orbital
separation, the eccentricity, which is initially $\sim 0.1$, will
decrease to $\sim 0.01$ within several orbits, because gravitational
radiation reaction has a strong effect to reduce the orbital
eccentricity \cite{PM}. However, the separation has to be large enough
for the large initial eccentricity to be reduced.  This implies that a
long-term simulation, which is not computationally favored, is
required. (As we illustrate in this paper, the eccentricity is not
sufficiently suppressed to $\alt 0.01$ in $\sim 5$ orbits if we adopt
the initial condition obtained in the $\beta^{\varphi}$ condition.)

In this paper, we employ a new condition in which the center of mass
is determined in a phenomenological manner: We impose the condition
that the total angular momentum of the system for a given value of
$m_0\Omega$ agrees with that derived by the 3PN approximation (see
Ref.~\cite{Husa} for a similar concept).  This can be achieved by
appropriately choosing a hypothetical position of the center of mass
(it may not be better to refer to this position as ``the center of
mass'' but simply as ``the rotation axis'').  With this method, the
total energy of the system does not agree completely with that derived
by the 3PN approximation \cite{KST}, and thus, the eccentricity does
not become zero. However, the resulting eccentricity in this condition
is much smaller than that in the $\beta^{\varphi}$ condition (see
Fig.~\ref{FIG1}), and thus, for a moderately long-term simulation (in
$\sim 2$--3 orbits), the effect of the eccentricity is suppressed to
an acceptable level for a scientific discussion, as shown in
Sec.~\ref{sec4}. We refer to this condition as ``3PN-J condition'' in
the following.

Figure \ref{FIG1} plots evolution of the orbital separation for ${\cal
C}=0.145$ and for $Q=2$ and 5. Here, the separation is defined as the
coordinate separation between the positions of the puncture and of the
maximum density of the NS, $r_{\rm sep}=|x^i_{\rm sep}|$; see
Eq. (\ref{eqsep}). For the results with the initial condition derived
in the $\beta^{\varphi}$ condition, the separation badly oscillates
with time and the amplitude of this oscillation is still conspicuous
even after 3--4 orbital motions. Consequently, the eccentricity is not
negligible even at the last orbit just before the merger. By contrast,
for the case that the initial condition is derived in the 3PN-J
condition, the amplitude of the oscillation is much smaller and,
furthermore, the amplitude of this oscillation does not become
conspicuous within about two orbital motions.  Although the
eccentricity does not become exactly zero, it is within an acceptable
level at the last orbit just before the merger (see, e.g.
gravitational waveforms shown in Sec. \ref{sec:gw}).

We note that the coordinate separation is a gauge-dependent quantity
and hence the discussion here is not based on the very physical
quantity. The physical quantity such as the orbital eccentricity is
not rigously extracted from it.  However, the similar quantitative
feature is seen if we plot evolution of the gravitational-wave
frequency as a function of time (which is the physical quantity); the
oscillation amplitude of the orbital separation shows the magnitude of
the orbital eccentricity at least approximately (see Sec. \ref{sec:gw} 
for more physical analysis).

\subsection{Chosen models} 

\begin{table*}[t]
\caption{Key parameters for the initial conditions adopted for the
numerical simulation in units of $\kappa=1$.  The mass ratio
($Q=M_{\rm BH}/M_{\rm NS}$), BH mass ($M_{\rm BH}$), mass parameter of
the puncture ($M_{\rm p}$), rest mass of the NS ($M_{*}$), mass
($M_{\rm NS}$) and compactness (${\cal C}=M_{\rm NS}/R_{\rm NS}$:
$R_{\rm NS}$ is the circumferential radius) of the NS when it is in
isolation, maximum density of the NS ($\rho_{\rm max}$), ADM mass of
the system ($M_0$), total angular momentum ($J_0$) in units of
$M_0^2$, orbital period ($P_0$) in units of $M_0$, and orbital angular
velocity ($\Omega_0$) in units of $m_0^{-1}$.  The first and second
numerical values described for the model name (the first column)
denote the values of $Q$ and ${\cal C}$, respectively. 
The first 10 models are computed in the 3PN-J condition and the 
last 4 are in the $\beta^{\varphi}$ condition.}
\begin{tabular}{cccccccccccc} \hline
3PN-J &&&&&&&&&&& \\ 
~~~Model~~~ 
& ~~$Q$~~ & ~~~$M_{\rm BH}$~~~ & ~~~$ M_{\rm p}$~~~ & ~~~~$M_{*}$~~~ 
& ~~~$M_{\rm NS}$~~~ & $M_{\rm NS}/R_{\rm NS}$ & ~~~$\rho_{\rm max}$~~~ 
& ~~~$M_0$~~~ & ~~$J_0/M_0^2$~~ & ~$P_0/M_0$~ & ~~$m_0\Omega_0$~~ \\ \hline
M15.145 & 1.5 & 0.2093 & 0.2043 & 0.1500 & 0.1395 & 0.145
& 0.1262 & 0.3452 & 0.8393 & 212.5 & 0.02987 \\ \hline
M20.145 &  2  & 0.2790 & 0.2736 & 0.1500 & 0.1395 & 0.145 
& 0.1261 & 0.4146 & 0.8603 & 213.7 & 0.02968 \\ \hline
M20.145N&  2  & 0.2790 & 0.2732 & 0.1500 & 0.1395 & 0.145 
& 0.1260 & 0.4143 & 0.8418 & 191.8 & 0.03309 \\ \hline
M20.160 &  2  & 0.2957 & 0.2900 & 0.1600 & 0.1478 & 0.160 
& 0.1512 & 0.4394 & 0.8608 & 214.4 & 0.02958 \\ \hline
M20.178 &  2  & 0.3119 & 0.3059 & 0.1700 & 0.1560 & 0.178 
& 0.1890 & 0.4635 & 0.8582 & 211.4 & 0.03001 \\ \hline
M30.145 &  3  & 0.4185 & 0.4121 & 0.1500 & 0.1395 & 0.145 
& 0.1255 & 0.5534 & 0.7091 & 192.2 & 0.03298 \\ \hline
M30.160 &  3  & 0.4435 & 0.4367 & 0.1600 & 0.1478 & 0.160 
& 0.1504 & 0.5864 & 0.7096 & 192.9 & 0.03285 \\ \hline
M30.178 &  3  & 0.4679 & 0.4608 & 0.1700 & 0.1560 & 0.178
& 0.1878 & 0.6187 & 0.7103 & 193.8 & 0.03269 \\ \hline
M40.145 &  4  & 0.5580 & 0.5512 & 0.1500 & 0.1395 & 0.145 
& 0.1252 & 0.6926 & 0.6042 & 192.1 & 0.03294 \\ \hline
M50.145 &  5  & 0.6975 & 0.6905 & 0.1500 & 0.1395 & 0.145 
& 0.1249 & 0.8318 & 0.5238 & 191.8 & 0.03296 \\ \hline
$\beta^{\varphi}$ &&&&&&&&&&& \\ 
Model & $Q$ & $M_{\rm BH}$ & $ M_{\rm p}$ & $M_{*}$ 
& $M_{\rm NS}$ & $M_{\rm NS}/R_{\rm NS}$ & $\rho_{\rm max}$ 
& $M_0$ & $J_0/M_0^2$ & $P_0/M_0$ & $m_0\Omega_0$ \\ \hline
M20.145b &  2 & 0.2790 & 0.2737 & 0.1500 & 0.1395 & 0.145 
& 0.1261 & 0.4144 & 0.8462 & 211.0 & 0.03007 \\ \hline
M30.145b &  3 & 0.4185 & 0.4121 & 0.1500 & 0.1395 & 0.145 
& 0.1256 & 0.5531 & 0.6960 & 189.5 & 0.03345 \\ \hline
M40.145b &  4 & 0.5580 & 0.5512 & 0.1500 & 0.1395 & 0.145 
& 0.1252 & 0.6922 & 0.5905 & 188.9 & 0.03351 \\ \hline
M50.145b &  5 & 0.6975 & 0.6905 & 0.1500 & 0.1395 & 0.145 
& 0.1250 & 0.8315 & 0.5134 & 189.1 & 0.03345 \\ \hline
\end{tabular}
\end{table*}

In the polytropic EOS, the adiabatic constant, $\kappa$, is a free
parameter, and thus, physical units such as mass, radius, and time can
be rescaled arbitrarily by simply changing the value if $\kappa$:
i.e., when a numerical result for a particular value (say
$\kappa=\kappa_1$) is obtained, we can also obtain the numerical
results of these quantities for $\kappa=\kappa_2$ by simply rescaling
them by a factor of $(\kappa_2/\kappa_1)^{1/2(\Gamma-1)}$.  This
implies that $\kappa$ can be completely scaled out of the problem.  In
this paper, we present the results in units of $\kappa=1$ (and
$c=G=1$), because such units are popular in other groups
\cite{ILLINOIS,CORNELL}. In the polytropic EOS, the nondimensional
quantities such as ${\cal C}=GM_{\rm NS}/c^2R_{\rm NS}$, $M_0\Omega$,
$M_0\kappa^{-1/2(\Gamma-1)}$, $R_{\rm NS}\kappa^{-1/2(\Gamma-1)}$, and
mass ratio ($Q$) are unchanged irrespective of the value of $\kappa$
and have an invariant meaning.

In the present work, the numerical simulation is performed for a wide
variety of initial conditions (see Table I), but for restricting that
the BH spin is zero.  We characterize the BH-NS binaries by the mass
ratio, $Q=M_{\rm BH}/M_{\rm NS}$, and the compactness of the NS,
${\cal C}=M_{\rm NS}/R_{\rm NS}$. (Note that in
Refs.~\cite{ST08,SACRA}, we use $q=1/Q$ instead of $Q$ to specify the
model.)  The mass ratio is chosen in the range between 1.5 and 5.0,
and the compactness of the NS is in the range, 0.145--0.178.  We note
that the typical mass of the NS in nature is 1.3--$1.4 M_{\rm NS}$
\cite{Stairs} and the likely lower bound of the BH mass is $\sim
2M_{\odot}$.  This implies that $Q$ should be chosen to be larger than
$\sim 1.5$.  The circumferential radius of an NS of $M_{\rm
NS}=1.35M_{\odot}$ and ${\cal C}=0.145$, 0.160, and 0.178 is 13.8,
12.5, and 11.2 km, which are reasonable values for modeling the NSs. 

\begin{figure}[t]
\epsfxsize=3.2in
\leavevmode
\epsffile{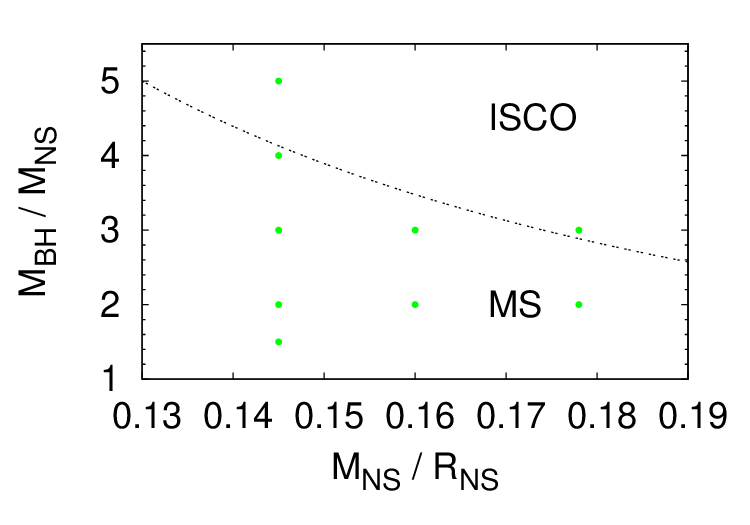}
\vspace{-5mm}
\caption{Initial conditions in the parameter space of $({\cal C}, Q)$
are plotted by the solid circles. The meaning of the dashed curve is
as follows \cite{TBFS2}: For the binaries located above the dashed
curve, mass-shedding (MS) does not occur until the binaries reach 
the ISCO, whereas for the binaries below the dashed curve, the
mass-shedding will occur for the NS due to the tidal force of the BH 
before the ISCO is reached. 
\label{FIG2}}
\end{figure}

Figure \ref{FIG2} shows the initial conditions in the parameter space
of $({\cal C}, Q)$. The meaning of the dashed curve, derived in
Ref.~\cite{TBFS2}, is as follows: For the binaries located above the
dashed curve, mass-shedding of the NS by the tidal effect of the
companion BH does not occur until the binary reaches the ISCO, whereas
for the binary shown below the dashed curve, the mass-shedding will
occur for the NS before the ISCO is reached and in such a system,
tidal disruption may subsequently occur. Thus, many of the NSs in the
chosen binary system will be subject to the tidal effect of the
companion BH in a close orbit, but for some of them (e.g., $({\cal C},
Q)= (0.145, 5)$ and (0.178, 3)), the tidal effect is unlikely to play
an essential role in the inspiral phase. We also note that an NS which
is in the mass-shedding is not always tidally disrupted immediately, 
although the mass-shedding NS is the candidate to be tidally disrupted. 
Namely, the condition for inducing the tidal disruption is in general
more restricted than that for inducing the mass-shedding (see Sec.~IV). 

In the present work, we prepare quasiequilibrium states with
$m_0\Omega_0 \approx 0.030$ for $Q=1.5$ and 2, and $m_0\Omega_0
\approx 0.033$ for $Q=2$--5, where $\Omega_0$ is the orbital angular
velocity. We choose these values of $\Omega_0$ so as for the BH-NS
binaries to experience more than 4 orbits before the onset of the merger.
In such a long-term evolution, the eccentricity which 
presents at $t=0$ decreases to an acceptable level during the inspiral
phase, and also the nonzero radial velocity associated with the
gravitational radiation reaction is approximately taken into account 
from the late inspiral phase.

In Table I, several key quantities for the quasiequilibrium states
adopted in this paper are listed.  Specifically, we prepare 10
models. The first two and last three numerical numbers described for
the model name denote the values of $Q$ and ${\cal C}$,
respectively. For comparison, the same quantities are also listed for
4 selected quasiequilibrium states obtained in the $\beta^{\varphi}$
condition. We find that the angular momentum of them is by $\sim 2\%$
smaller than that of the quasiequilibrium states obtained in the 3PN-J
condition for $m_0 \Omega_0 \approx 0.033$.

\section{Preparation for numerical simulation}

\begin{table*}[tbh]
\caption{Parameters of the grid structure for the numerical simulation
with our AMR algorithm.  In the column named ``Levels'', the number of
total refinement levels is shown (in the bracket, the numbers of
coarser and finer levels are specified; see Ref.~\cite{SACRA} for the
definition of the coarser and finer levels).  $\Delta x(=h_7)$ is the
minimum grid spacing for $N=36$, $R_{\rm diam}$ the coordinate length
of semi-major diameter of the NS, $L$ the location of outer boundaries
along each axis, $\lambda_0(=\pi/\Omega_0)$ the gravitational
wavelength at $t=0$, and $\Delta x_{\rm gw}$ the grid spacing, by
which gravitational waves are extracted for $N=36$.  Note that $M_{\rm
BH}$ denotes the BH mass (the irreducible mass) at $t=0$.  The grid
structures for models M20.145b--M50.145b are the same as those for
models M20.145--M50.145, respectively. For the simulations with
$N\not=36$, the size of each domain (and hence $L$) is unchanged, and
thus, the grid spacing $h_l$ is simply changed.
\label{BHNSGRID}}
\begin{tabular}{cccccc} \hline
Run & ~~Levels~~ & $\Delta x/M_0~(\Delta x/M_{\rm BH})$ &
~$R_{\rm diam}/\Delta x$~
& ~$L/M_0~(L/\lambda_0)$~ & ~$\Delta x_{\rm gw}/M_0$~ \\ \hline
M15.145 & 8~(4+4) &0.0407~(0.0672)& 112  & 187.7~(1.77)&1.30--5.21\\ \hline
M20.145 & 8~(4+4) &0.0377~(0.0560)& 99.3 & 173.7~(1.63)&1.21--4.82\\ \hline
M20.145N& 8~(4+4) &0.0377~(0.0560)& 99.4 & 173.8~(1.81)&1.21--4.83\\ \hline
M20.160 & 8~(4+4) &0.0356~(0.0528)& 92.9 & 163.9~(1.53)&1.14--4.55\\ \hline 
M20.178 & 8~(4+4) &0.0324~(0.0481)& 89.1 & 149.1~(1.41)&1.04--4.14\\ \hline 
M30.145 & 8~(4+4) &0.0305~(0.0403)& 90.0 & 140.5~(1.46)&0.98--3.90\\ \hline
M30.160 & 8~(3+5) &0.0266~(0.0352)& 91.2 & 122.8~(1.27)&0.85--3.41\\ \hline
M30.178 & 8~(3+5) &0.0253~(0.0334)& 84.1 & 116.3~(1.20)&0.81--3.23\\ \hline
M40.145 & 8~(3+5) &0.0244~(0.0302)& 89.0 & 112.3~(1.17)&0.78--3.12\\ \hline
M50.145 & 8~(3+5) &0.0203~(0.0242)& 88.4 & 93.5~(0.98) &0.65--2.60\\ \hline
\end{tabular}
\end{table*}

\subsection{Brief summary of formulation and methods}

The numerical simulations are performed using a code {\tt SACRA}
recently developed in our group \cite{SACRA}.  The details of the chosen 
scheme, formulation, gauge condition, and methods for the analysis 
are described in Ref.~\cite{SACRA} to which the reader may refer.
Reference \cite{SACRA} also shows that {\tt SACRA} can successfully
simulate the inspiral and merger of BH-BH, NS-NS, and BH-NS binaries. 

In {\tt SACRA}, the Einstein equations are solved in a moving-puncture
version \cite{BB,BB2,BB4} of the Baumgarte-Shapiro-Shibata-Nakamura
formalism \cite{BSSN}.  Specifically, we evolve $W \equiv
\gamma^{-1/6}$ \cite{WWW}, conformal three-metric, $\tilde \gamma_{ij}
\equiv \gamma^{-1/3}\gamma_{ij}$, the tracefree extrinsic curvature,
$\tilde A_{ij} \equiv \gamma^{-1/3}(K_{ij}-K\gamma_{ij}/3)$, the trace
part of $K_{ij}$, $K$, and a three auxiliary variable, $\Gamma^i
\equiv -\pa_j \tilde \gamma^{ij}$ (or $F_i \equiv \pa_j \tilde
\gamma_{ij}$).  Here, $\gamma_{ij}$ is the three-metric, $K_{ij}$ the
extrinsic curvature, and $\gamma \equiv {\rm det}(\gamma_{ij})$.  In
the numerical simulation, a fourth-order finite differencing scheme in
space and time is used implementing an adaptive mesh refinement (AMR)
algorithm (at refinement boundaries, a second-order interpolation
scheme is partly adopted).  The advection term such as
$\beta^i\partial_i \tilde \gamma_{jk}$ is evaluated by a fourth-order
non-centered finite-difference, as proposed in Ref.~\cite{BB4}.

The moving-puncture approach is adopted for evolving the BH 
\cite{BB2} (see also Ref.~\cite{BB3}); 
i.e., we adopt one of the moving-puncture gauge 
conditions as \cite{BB4} 
\beqn
&&(\pa_t -\beta^j\pa_j) \beta^i=0.75B^i, \\
&&(\pa_t -\beta^j\pa_j) B^i =(\pa_t -\beta^j\pa_j) \tilde \Gamma^i
-\eta_s B^i,
\label{shift2} 
\eeqn
where $B^i$ is an auxiliary variable and $\eta_s$ is an arbitrary
constant. In the present paper, we set $\eta_s=0.414$ in units of 
$\kappa=c=G=1$. 

For the numerical hydrodynamics, we evolve $\rho_* \equiv \rho \alpha
u^t e^{6\phi}$, $\hat u_i \equiv h u_i$, and $e_* \equiv \rho h \alpha
u^t -P/(\rho \alpha u^t)$.  To handle the advection terms in the
hydrodynamic equations, a high-resolution central scheme \cite{KT} is
adopted with a third-order piecewise parabolic interpolation and with
a steep min-mod limiter in which the limiter parameter $b$ is set to
be 3 (see appendix A of Ref.~\cite{S03}).  We adopt the $\Gamma$-law
EOS in the simulation as 
\beq 
P=(\Gamma-1)\rho \varepsilon, 
\eeq 
where $\Gamma=2$.

Properties of the BHs such as mass and spin are determined by
analyzing area and circumferential radii of apparent horizons. 
A numerical scheme of our apparent horizon finder is described in 
Refs. \cite{SACRA,AH}. 

Gravitational waves are computed by extracting the outgoing part of
the Newman-Penrose quantity (the so-called $\Psi_4$).  The extraction
of $\Psi_4$ is carried out for several constant coordinate radii, $r 
\approx 50$--$100M_0$. The plus and cross modes of gravitational waves
are obtained by performing time integration of $\Psi_4$ twice, with
appropriate choice of integration constants and subtraction of
unphysical drift which is caused primarily by the drift of the center
of mass of the system.  (Because we extract $\Psi_4$ for fixed, finite
coordinate radii, the drift of the center of mass spuriously affects 
gravitational waveforms.) Specifically, whenever the time integration
is performed, we subtract a function of the form $a_2 t^2 + a_1 t +
a_0$ where $a_0$--$a_2$ denote constants which are determined by the
least-square fitting of the numerical data.

We compute the modes of $2 \leq l \leq 4$ and $|m| \leq l$ for
$\Psi_4$, and found that the quadrupole mode of $(l, |m|)=(2, 2)$ is
always dominant, but $(l, |m|)=(3, 3)$, (4, 4), and (2, 1) modes also
contribute to the energy and angular momentum dissipation by more than
$1\%$ for some of models (in particular for large values of $Q$;
cf. Table IV). 

We also estimate the kick velocity from the linear momentum flux of
gravitational waves. The linear momentum flux $dP_i/dt$ is computed by
the same method as that given in, e.g., Refs.~\cite{BB4,kick2}.
Specifically, the coupling terms between $(l, m)=(2, \pm 2)$ and $(3,
\mp 3)$ modes, between $(2, \pm 2)$ and $(2, \mp 1)$ modes, and
between $(3, \pm 3)$ and $(4, \mp 4)$ modes contribute primarily to 
the linear momentum flux \cite{Buonanno}. From the total linear
momentum dissipated by gravitational waves, \beqn \Delta P_i = \int
{dP_i \over dt} dt, \eeqn the kick velocity is defined by $\Delta
P_i/M_0$ where $M_0$ is the initial ADM mass of the system.

\subsection{Setting grids for AMR scheme}

The Einstein and hydrodynamic equations are solved in an AMR algorithm
described in Ref.~\cite{SACRA}. In the present work, we prepare 8
refinement-level domains of different grid resolutions and domain
sizes.  Each domain is composed of the uniform vertex-center-grid with
grid number $(2N+1,2N+1,N+1)$ for $(x,y,z)$ where $N$ is a constant
and is always chosen to be 24, 30, and 36 to check convergence of
numerical results.  The equatorial plane symmetry with respect to the
$z=0$ plane is assumed. The length of a side for the largest domain is
denoted by $2L$, and the grid spacing for each domain is $h_l=L/N/2^l$
for $l=0$--7. As described in Ref.~\cite{SACRA}, the regions around a BH
and an NS are covered by ``finer'' domains which move with the BH and
NS. On the other hand, ``coarser'' domains which cover wider regions 
do not move and their center is fixed approximately to be the 
center of mass of the system. 

Table II lists the parameters for the grid structure in our AMR
scheme.  $\lambda_0$ denotes the wavelength of gravitational waves at
$t=0$ ($\lambda_0=\pi/\Omega_0$). For all the cases, $L$ is chosen to
be 1--2$\lambda_0$, implying that the outer boundaries are located in
a wave zone. The NS is covered by the finest and second-finest domains
and the coordinate radius of the apparent horizon for the BH is always
covered by more than 15 grid points for $N=36$. Because the numerical
results (such as the time spent in the inspiral phase, the mass and
spin for the final state of the BH, and total energy radiated by
gravitational waves) for $N \geq 30$ depend only weakly on the grid
resolution, we conclude that the convergence is approximately achieved
for $N=36$ (except for the rest mass of disks which are formed for the
model with small values of ${\cal C}$ and $Q$; see Sec. \ref{sec4.1}).

For $N=36$, the total memory required for the simulation with 13
domains is about 5 GBytes. We perform all the simulations using
personal computers of 8 GBytes memory and 2--8 processors (in one job,
only two processors are used with an OpenMP library).  The typical
computation time for one model with $N=36$ is 5--7 weeks on the
personal computers of clock speed 3 GHz. 

\begin{figure*}[t]
\epsfxsize=2.7in
\leavevmode
(a)\hspace{-1cm}\epsffile{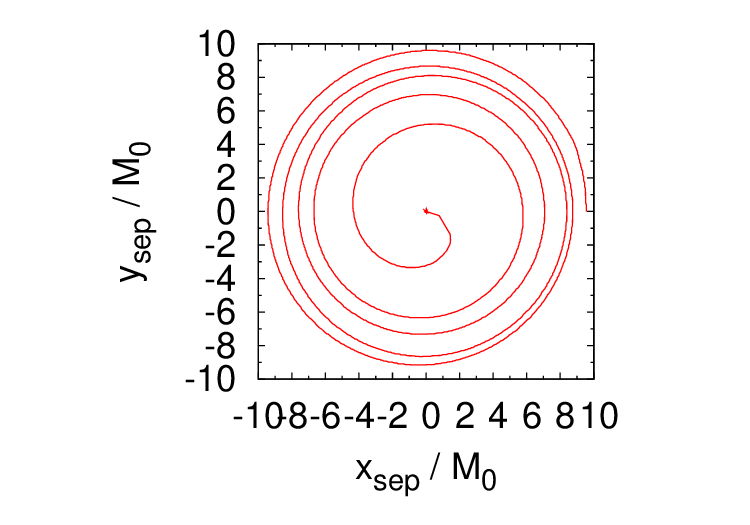}
\epsfxsize=2.7in
\leavevmode
\hspace{-0.5cm}(b)\hspace{-1cm}\epsffile{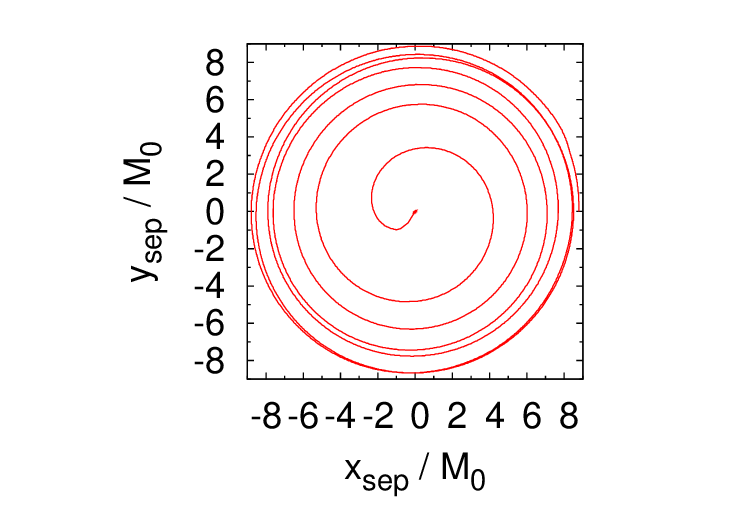}
\epsfxsize=2.7in
\leavevmode
\hspace{-0.5cm}(c)\hspace{-1cm}\epsffile{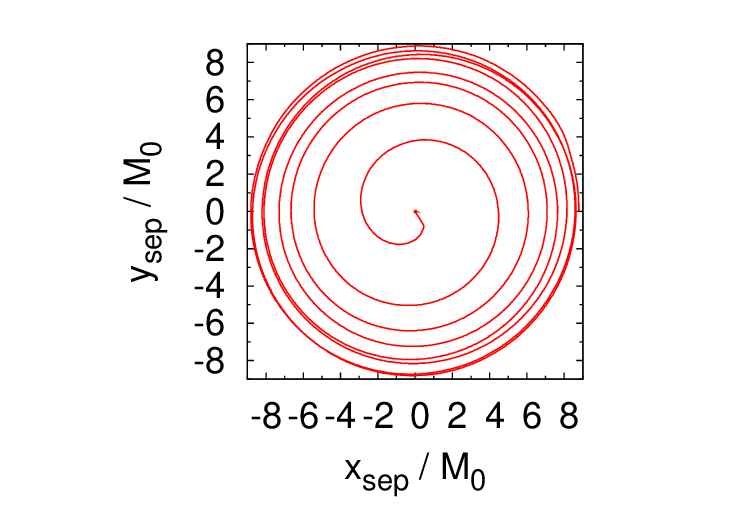}
\vspace{-2mm}
\caption{Coordinate separation between the BH and NS (a) for models
M20.145, (b) M40.145, and (c) M50.145. Here, $r_{\rm sep}=|x^i_{\rm
NS}-x^i_{\rm BH}|$; see Eq.~(\ref{eqsep}).
\label{FIG3}}
\end{figure*}

\subsection{Atmosphere}

Because any conservation scheme of hydrodynamics is unable to evolve a
vacuum, we have to introduce an artificial atmosphere outside the NSs.
The density of the atmosphere has to be small enough to exclude its
spurious effect to the orbital motion of the BH and NS, and to avoid
overestimation of the total rest mass for a disk surrounding the BH,
which may be formed after the merger if the NS is tidal disrupted. We
initially assign a small rest-mass density for the atmosphere as follows 
\beqn 
\rho_{\rm atmo}=\left\{
\begin{array}{ll}
\rho_{\rm crit} & r \leq r_0 \\
\rho_{\rm crit}e^{1-r/r_0} & r > r_0.
\end{array}
\right.
\eeqn 
where $r_0$ is a constant chosen to be $5L/32$.  We choose $\rho_{\rm
crit}=\rho_{\rm max} \times 10^{-9}$ where $\rho_{\rm max}$ is the
maximum rest-mass density of the NS initially given.  With such
choice, the total amount of the rest mass in the atmosphere is less
than $10^{-5}M_*$. Accretion of the atmosphere of such small mass onto
the BH and NS plays a negligible role for their orbital evolution. As
we state in Sec. I, one of the important tasks for the merger
simulation of the BH-NS binaries is to determine the rest mass of
disks surrounding a BH formed after the merger. In the simulation, we
pay attention only to the case that the disk mass is larger than
$10^{-5}M_*$. 

During the evolution, we also adopt an artificial treatment for the
low-density region in the following manner: (i) If the density is
smaller than $\rho_{\rm atmo}$, we set $\rho=\rho_{\rm atmo}$ and
$u_i=0$. Then, the specific internal energy $\varep$ is set to be
$\kappa \rho_{\rm crit}^{\Gamma-1}/(\Gamma-1)$. (ii) Even if the
density is larger than $\rho_{\rm atmo}$, we reduce the specific
momentum $h u_i$ by a factor of $1-\exp[-\rho/\rho_{\rm crit}]$; i.e.,
for a fluid of density $\alt 5 \rho_{\rm crit}$, the specific momentum
is artificially reduced. The reason that we adopt these treatments is
that a numerical instability resulting in a negative density or
pressure often happens accidentally for the low-density region in the
absence of the artificial treatment. However, by limiting the
unphysical growth of the specific momentum, as described above, such
instabilities are excluded. For some models, we checked whether the
magnitude of $\rho_{\rm crit}$ affects the numerical result, but as
long as the small value is chosen for it as in the present work,
dependence of the numerical results on the magnitude of $\rho_{\rm
crit}$ is quite tiny.

\section{Numerical results}\label{sec4}

\begin{figure*}[th]
\caption{Snapshots of the density contour curves and density contrasts
as well as the location of the BH in the merger and ringdown phases for
model M20.145.  The contour curves are plotted for $\rho=10^{-i}$
where $i=2, 3, 4, 5$ in the first four panels, whereas in the
last two, $\rho=10^{-i}$ where $i=3$, 4, and 5.  The first panel
denotes the state just at the onset of the merger. The filled circles
show the region inside the apparent horizon. Note that the maximum value
of $\rho$ for the NS in the inspiral phase is $\approx 0.126$. 
\label{FIG4}}
\end{figure*}

\subsection{Orbital evolution, tidal disruption, and disk mass}\label{sec4.1}

Figure \ref{FIG3} plots evolution of the coordinate separation 
between the BH and NS, $x^i_{\rm sep}$, for models M20.145, 
M40.145, and M50.145. $x^i_{\rm sep}$ is defined by 
\beq 
x^i_{\rm sep}=x^i_{\rm NS}-x^i_{\rm BH}, \label{eqsep}
\eeq 
where $x^i_{\rm NS}$ and $x^i_{\rm BH}$ denote the positions of the
maximum rest-mass density of the NS and of the puncture,
respectively. This figure illustrates that the binaries are 
in a slightly eccentric orbit for the first $\sim 2$ orbits because the 
strictly circular orbit is not provided initially. Also, 
in the first $\sim 2$ orbits, decrease rate of the orbital
separation due to gravitational radiation reaction is not as large as
that predicted by the PN theory, in particular for larger
values of $Q$.  However, because the eccentricity decreases due to
the gravitational radiation reaction during the evolution and also the
initial eccentricity is not very large (see, e.g., Fig.~\ref{FIG1}),
the orbit approaches approximately to a quasicircular orbit
after a few orbits. The resulting final 2--3 orbits before the onset of 
merger appear to be close to a quasicircular one. This behavior is
much better than that obtained when an initial condition computed in
the $\beta^{\varphi}$ condition is adopted 
(see, e.g. Fig.~15 of Ref.~\cite{SACRA}). 

For models M20.145, M40.145, and M50.145, the binaries spend in the
inspiral phase for $\sim 4.8$, 6.5, and 7.8 orbits, respectively. For
the larger values of $Q$ with an approximately fixed value of
$m_0\Omega_0$, the number of the inspiral orbit is larger, because the
luminosity of gravitational waves is approximately proportional to
$Q^2/(1+Q)^4$, and as a result, the binary evolves more slowly for the
larger values of $Q$.


\begin{figure*}[p]
\caption{The same as Fig.~\ref{FIG4} but for model M40.145. 
\label{FIG5}}
\end{figure*}

\begin{figure*}[p]
\caption{The same as Fig.~\ref{FIG4} but for model M50.145. 
\label{FIG6}}
\end{figure*}

\begin{figure*}[t]
\epsfxsize=3.2in
\leavevmode
\epsffile{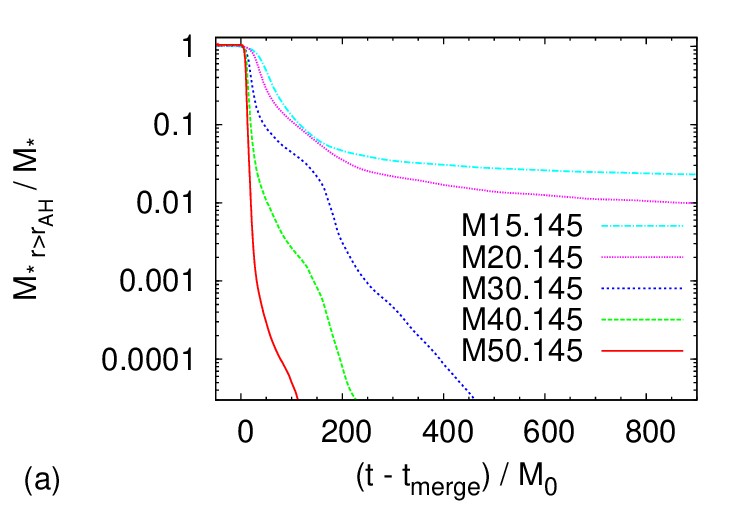}
\epsfxsize=3.2in
\leavevmode
~~~~\epsffile{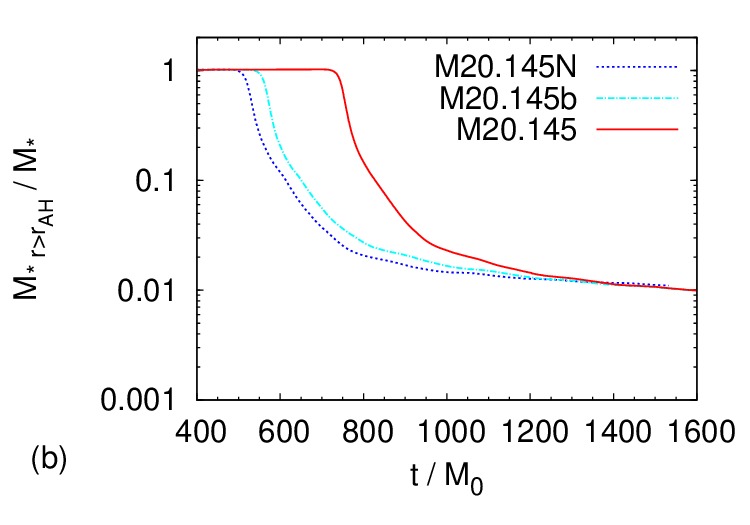}\\
\epsfxsize=3.2in
\leavevmode
\epsffile{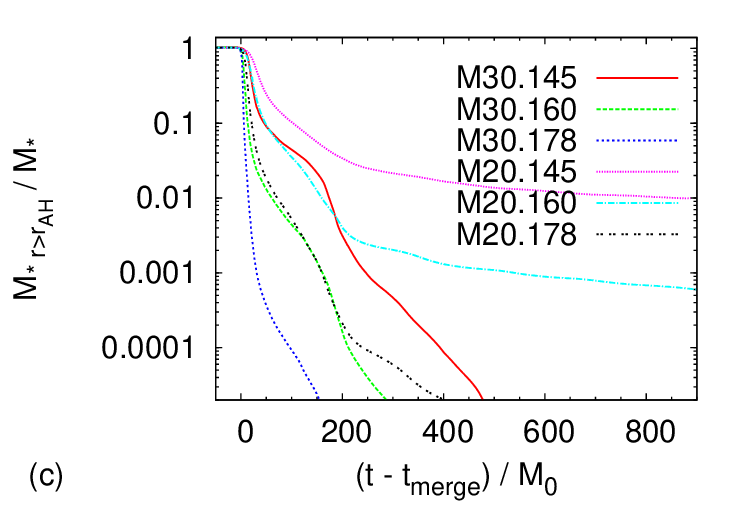}
\epsfxsize=3.2in
\leavevmode
~~~~\epsffile{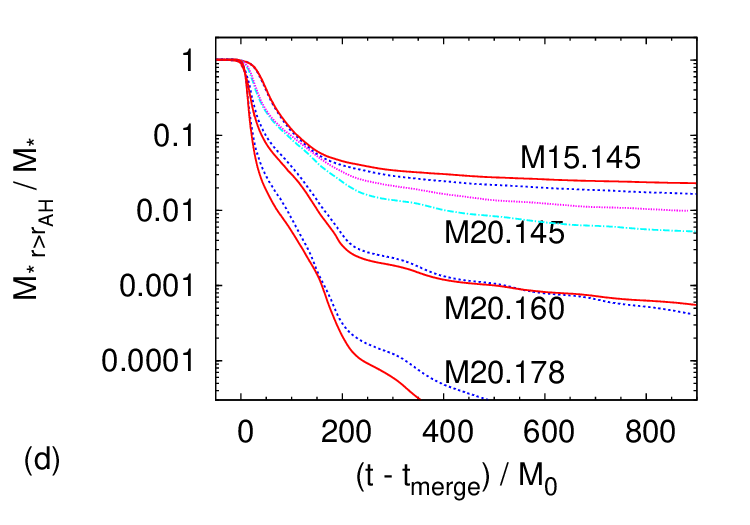}
\vspace{-4mm}
\caption{Evolution of the rest mass of the material located outside
the apparent horizon (a) for models M15.145, M20.145, M30.145,
M40.145, and M50.145 with $N=36$, (b) for model M20.145, M20.145N, and
M20.145b with $N=36$, (c) for models M20.145, M20.160, M20.178,
M30.145, M30.160, and M30.178 with $N=36$, and (d) for models M15.145,
M20.145, M20.160, and M20.178 with $N=36$ (solid curves) and $N=30$
(dotted curves).  For a hypothetical value of $M_{\rm NS}=1.35M_{\odot}$, 
$100M_0 \approx 1.66$, 2.00, 2.66, 3.33, and 3.99 ms for 
$Q=1.5$, 2, 3, 4, and 5, respectively.
\label{FIG7}}
\end{figure*}

Figures \ref{FIG4}--\ref{FIG6} plot late-time evolution of the
rest-mass density contour curves and density contrasts as well as the
location of the BHs for models M20.145, M40.145, and M50.145,
respectively. For model M20.145, the NS is tidally disrupted in the
late inspiral phase (see the first panel of Fig.~\ref{FIG4}).
Subsequently the material of the NS forms a one-armed spiral arm
around its companion BH (second panel of Fig.~\ref{FIG4}). In the
spiral arm, a transport process of angular momentum from its inner to
outer region is likely to work efficiently.  Then, the spiral arm
winds around the BH (third panel of Fig.~\ref{FIG4}), and most of the
fluid elements, which do not have angular momentum large enough to
have an orbit around the BH, fall into the BH. In the first $\sim 200
M_0$ after the tidal disruption, $\sim 98\%$ of the material falls
into the BH (see Fig.~\ref{FIG7}). However, a small fraction of the
fluid elements obtain angular momentum large enough to escape the
capture by the BH, and form a disk around the BH. For $t-t_{\rm
merger} \agt 300 M_0$, where $t_{\rm merger}$ approximately denotes
the time at which the merger sets in, accretion rate decreases, and
then, the disk relaxes to a quasisteady state (fourth--sixth panels).
For model M20.145, the rest mass of the disk is $\sim 0.01M_*$ at
$t-t_{\rm merger}\approx 1000M_0$ for $N=36$
(cf. Fig.~\ref{FIG7}). For a hypothetical value of $M_{\rm
NS}=1.35M_{\odot}$, $1000M_0$ is approximately equal to 20 ms, and
thus, the lifetime of the formed disk is likely to be much longer than
20 ms. Also, for a hypothetical value of $M_{\rm NS}=1.35M_{\odot}$,
$\rho=10^{-4}$ in the units of $c=G=\kappa=1$ corresponds to 
$\rho\approx 6.0 \times 10^{11}~{\rm g/cm^3}$. Thus, for that
hypothetical mass, the rest-mass density of the disk is high as $\sim
10^{11}$--$10^{12}~{\rm g/cm^3}$. Evolution and the final outcome for
model M15.145 are similar to those for model M20.145, although 
the disk mass is by a factor of $\sim 2$ larger due to the smaller 
value of $Q$.  

The NS for model M40.145 is also subject to tidal deformation and mass
shedding by the tidal effects of the companion BH in the late inspiral
phase, as indicated in Fig.~\ref{FIG2} and in the first panel of
Fig.~\ref{FIG5}. However, the tidal effects in this binary become
important only for the inspiral orbits close to the ISCO. Because the
approaching velocity is a substantial fraction of the orbital velocity
at such close orbits, the time scale available for the NS to be
tidally deformed before the onset of the merger is too short to
efficiently transport angular momentum inside the NS. As a result,
one-armed spiral arm is formed in a less conspicuous manner than that
for model M20.145 (see the second panel of Fig.~\ref{FIG5}), although
the NS is highly elongated at the merger.  Rather, most of the
material of the NS falls into the BH in a short time scale $\sim 200M_0$
(see the third panel of Fig.~\ref{FIG5}). A tiny fraction of the
material still spreads outward during the merger phase (see the fourth
panel of Fig.~\ref{FIG5}), but specific angular momentum for such
material is not large enough to escape the capture by the BH. In this
case, more than 99.999\% of the material is eventually swallowed by
the BH (cf. Fig.~\ref{FIG7}).

For model M50.145, even the mass shedding does not occur in the
inspiral phase, as indicated in Fig.~\ref{FIG2}. During the merger,
the NS is deformed by the tidal field of its companion BH (see the
first and second panels of Fig.~\ref{FIG6}), but spiral arm is not
formed nor angular momentum transport work. As a result, nearly all
the materials are swallowed by the BH in a short time scale of $\alt
100M_0$ (see the third and fourth panels of Fig.~\ref{FIG6}), and the
final outcome is a rotating BH approximately in a vacuum spacetime.

To clarify the infall process of the material into the 
companion BH, we plot evolution of the rest mass of the material 
located outside the apparent horizon, $M_{r>r_{\rm AH}}$, for several 
models in Fig.~\ref{FIG7}. Here, $M_{r> r_{\rm AH}}$ is defined by
\beqn
M_{r> r_{\rm AH}} \equiv \int_{r > r_{\rm AH}} \rho_* d^3x,
\eeqn
and $r_{\rm AH}(\theta, \varphi)$ denotes the radius of the apparent 
horizon for given angular coordinates. 

Figure \ref{FIG7}(a) plots $M_{r > r_{\rm AH}}$ as a function of
$t-t_{\rm merge}$, where $t_{\rm merge}$ denotes the approximate onset
time of the merger, for models M15.145, M20.145, M30.145, M40.145, and
M50.145.  This shows that (i) for $Q \leq 2$, a disk of rest mass
$\sim 0.01$-- $0.02M_*$ is formed and the lifetime of the disk is much
longer than the dynamical time scale; (ii) for $Q \geq 3$,
approximately all the materials are swallowed by the BH in $\sim
500M_0$ (the rest mass of the material located outside the apparent
horizon is less than $10^{-5}M_*$ at the final stage; see also Table
III). The time scale for the infalling is shorter for the larger values of
$Q$. These facts hold irrespective of the initial orbital separation
and grid resolution, as long as $Q \geq 3$.  Remember the fact that
the compactness of the NSs for these models is 0.145, which is a
relatively small value; the typical compactness for the NS of mass
1.3--$1.4M_{\odot}$ is between $\sim 0.14$ and $\sim 0.21$ according
to theories of high-density matter \cite{LP}.  Because the less
compact NS (i.e., the NS of larger radius) is more subject to disk
formation, we conclude that the disk (or torus) mass around a BH
formed after the merger is negligible, for the mass ratio $Q \geq
3$. By contrast, if the mass ratio is smaller than $\sim 2$ and the
compactness of the NS is relatively small as $\cC=0.145$, a disk of
mass $\sim 0.01$--$0.02M_*$ may be formed. (We note that in the
presence of BH spin, this conclusion changes; work in progress in our
group.)

To show the further evidence that the disk is really formed for $Q=2$
and ${\cal C}=0.145$, we generate Fig.~\ref{FIG7}(b).  In this figure,
we plot $M_{r > r_{\rm AH}}$ as a function of time for model M20.145,
for model M20.145N for which the initial orbital separation is smaller
than that of model M20.145, and for model M20.145b for which the
initial condition is computed in the $\beta^{\varphi}$ condition. This
figure shows that the disk mass at $t-t_{\rm merger} \approx 1000M_0$
is $\sim 0.01M_{*}$ irrespective of the initial conditions for a given
grid resolution ($N=36$).  This fact indicates that the initial
orbital separation is large enough to exclude spurious effects
associated with noncircularity of the initial condition to the
resulting disk mass. We note that in the previous early-stage works
\cite{ST08,ILLINOIS}, the simulations were performed from the initial
conditions of small orbital separations, and consequently, it was
found that the disk mass depends strongly on the initial separation,
failing to draw the definitive conclusion about the disk mass. This
drawback is overcome in this work.

To show dependence of $M_{r > r_{\rm AH}}$ as a function of $t$ on the
compactness of the NSs, in Fig.~\ref{FIG7}(c), we compare the results
for models M20.145, M20.160, and M20.178, and for models M30.145,
M30.160, and M30.178. As this figure shows, the disk mass
systematically and steeply decreases with the increase of the
compactness ${\cal C}$ for $Q=2$; the disk mass for model M20.160 is by
a factor of $\sim 15$ smaller than that for model M20.145. This
dependence is simply caused by the fact that the tidal disruption of
more compact NSs occurs for an orbit closer to the ISCO, suppressing
disk formation. This result implies that even for a small mass ratio
$Q=2$, disks are not formed if the radius of the NS is not very
large. Note that for a hypothetical mass of $M_{\rm NS}=1.35M_{\odot}$
for model M20.160, the circumferential radius of the NS is $R_{\rm
NS}=12.5$ km, which is not a large value because nuclear theories
predict it in the range $\sim 10$--15 km \cite{LP}.  Nevertheless, the
disk is not formed. This implies that for a typical NS of
$M_0=1.35M_{\odot}$ and $R_{\rm NS}=11$--12 km, the disk is formed
only for a highly restricted case, $Q < 2$, i.e., $M_{\rm BH} <
2.7M_{\odot}$. This conclusion agrees with a conjecture by Miller
\cite{CMiller}, and is also consistent with the results by Duez et
al. \cite{CORNELL} in which they show that the disk mass is at most
$\sim 0.01M_*$ for a compact NS of ${\cal C}=0.174$ with the most
optimistic mass ratio $Q=1$.  The present result also suggests that
most of BH-NS binaries may not be a promising candidate for the
central engine of SGRBs \cite{GRBdisk,GRBdisk1}, unless the radius of
the NS is fairly large $\agt 14$ km or the BH has a spin.

Figure \ref{FIG7}(c) also compares $M_{r > r_{\rm AH}}$ as a function
of $t$ for models M30.145, M30.160, and M30.178. For all these cases,
approximately all the materials of the NS eventually fall into the BH
in $\sim 500M_0$ after the onset of the merger. However, the merger
process and subsequent infalling process into the BH depend strongly
on the compactness of the NSs. For model M30.145, the mass-shedding
and subsequent tidal disruption occur before the binary reaches the
ISCO, as in model M20.145 (see also Fig.~18 of Ref.~\cite{SACRA} for
which essentially the same result as for model M30.145 is shown). As a
result, a spiral arm is formed around the companion BH and a fraction
of the material spreads outward. Then, a disk of rest mass $\agt
0.1M_*$ surrounding the BH is transiently formed, although a large
fraction of the material is swallowed by the BH in $\sim 50M_0$ after
the onset of the merger.  Subsequently, the material gradually
accretes from the disk into the BH in the time duration of $\sim
150M_0$.  During this phase, the disk mass decreases from $\sim
0.1M_*$ to $\sim 0.02M_*$, and thus, for the first $\sim 150M_0$ after
its formation, a massive disk is present.  However, the disk material
does not have sufficiently large specific angular momentum for keeping
orbits around the BH, and eventually falls into the BH in a runaway
manner.  In the end, more than 99.999\% of the material is swallowed
by the BH.

For model M30.160, a disk is formed around the BH transiently.
However, its mass is much smaller than that for model M30.145 because
the tidal disruption occurs at an orbit close to the ISCO, as in model
M40.145.  For model M30.178, a disk is not formed around the BH even
transiently, as in model M50.145.  The reason is that the NS in this
model is not tidally disrupted before the binary reaches the ISCO.
For these two cases, more than 99.999\% of the material is swallowed
into the BH within a short duration of 200--$300M_0$.

Before closing this section, we note that the final disk mass depends
very weakly on the grid resolution for models M30.145, M30.160,
M30.178, M20.160, and M20.178, whereas for models M15.145 and M20.145
for which the disk mass is $\agt 0.01M_*$, the final disk mass {\em
increases} with improving the grid resolution [see
Fig.~\ref{FIG7}(d)].  This implies that (i) for the case that the
massive disk is not formed, our conclusion is based on the convergent
result, whereas (ii) for the case that a disk of mass $\agt 0.01M_*$
is formed, the results with $N=36$ should be regarded as a lower-bound
for the disk mass, and the disk mass would be larger than $0.01M_*$
for model M20.145 and $0.02M_*$ for model M15.145. This dependence on
the grid resolution results from the fact that with the poorer grid
resolutions, numerical dissipation of angular momentum is larger,
increasing an amount of the material which falls into the BH. However,
this systematic dependence shows that the disk of mass $\agt 0.01M_*$
is indeed formed.

\subsection{Black hole mass and spin after merger}\label{BHspin}

\begin{figure}[t]
\begin{center}
\epsfxsize=3.1in
\leavevmode
(a)\epsffile{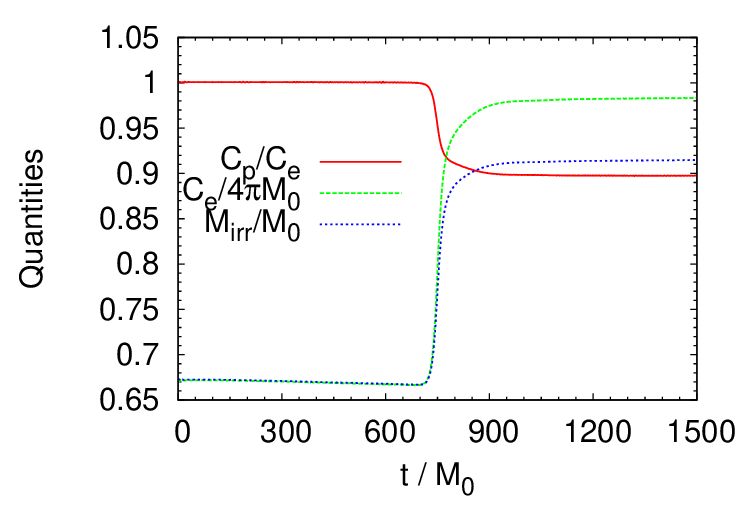}\\
\epsfxsize=3.1in
\leavevmode
(b)\epsffile{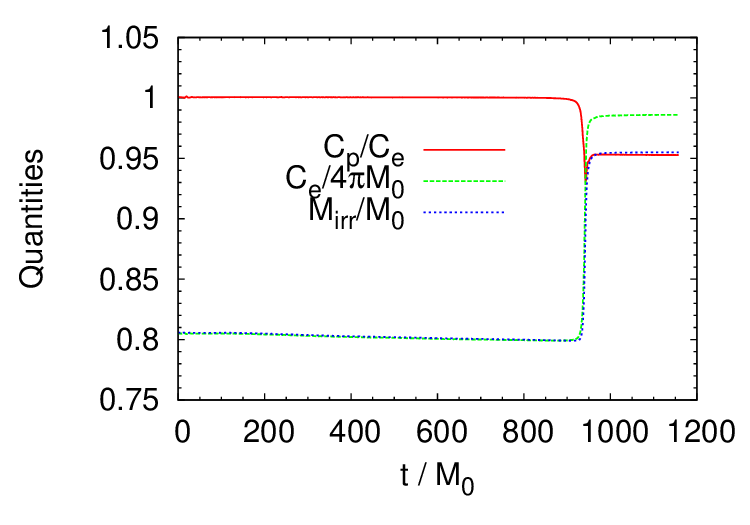}
\end{center}
\vspace{-3mm}
\caption{Evolution of $C_p/C_e$, $C_e/4\pi M_0$, and $M_{\rm irr}/M_0$ as 
functions of time (a) for models M20.145 and (b) M40.145. 
\label{FIG8}}
\end{figure}

Figure \ref{FIG8} plots $M_{\rm irr}/M_0$, $C_p/C_e$, and $C_e/4\pi
M_0$ of a BH as functions of time for models M20.145 and M40.145
(similar behavior is found for other models). Here, $C_p$ and $C_e$
are polar and equatorial circumferential radii of the BH.  An
irreducible mass, $M_{\rm irr}$, of the BH is defined by the area of
apparent horizon, $A_{\rm AH}$,
\beqn
M_{\rm irr}=\sqrt{{A_{\rm AH} \over 16\pi}}. 
\eeqn
$C_e/4\pi$ is equal to the BH mass in stationary vacuum spacetimes of
a BH. We follow its evolution, assuming that it is approximately 
equal to the BH mass even in the dynamical spacetime. 

Figure \ref{FIG8} shows that the values of these three quantities
remain approximately constant before the onset of the merger (more
specifically, before the material of the NS falls into the companion
BH). Because the BH is not spinning initially, the hypothetical ``BH
mass'', $C_e/4\pi$, should be approximately equal to the irreducible
mass $M_{\rm irr}$, and $C_p/C_e$ is approximately equal to unity.
These hold except for small numerical error of magnitude $\sim 1\%$.
After the onset of tidal disruption, $C_e/4\pi$ and $M_{\rm irr}$
quickly increase as the material of the NS falls into the BH, and
finally, they approximately reach constants.  By contrast, $C_p/C_e$
decreases due to spin-up of the BH caused by mass accretion.  Because
of the presence of the BH spin, $C_e/4\pi$ becomes unequal to the
irreducible mass after the onset of the merger.

In addition to $C_e/4\pi$, the mass of a BH formed after merger may be
estimated approximately by evaluating the total energy dissipated by
gravitational waves, $\Delta E$, and the baryon rest mass of disks 
surrounding the BH from an approximate relation of energy conservation as
\beqn
M_{\rm BH,f} \equiv M_0 - M_{r> r_{\rm AH}}-\Delta E. \label{mass_bhf}
\eeqn
We note that in this formula, we ignore the binding energy
between the BH and surrounding material. Thus, $M_{\rm BH,f}$ is likely 
to give a slightly overestimated value for the true BH mass.  

The values of $\Delta E$, $M_{r> r_{\rm AH}}$, $M_{\rm BH,f}$, and
$C_e/4\pi$ are listed for all the models chosen in this paper in Table
\ref{TABLE3} and IV. In Table \ref{TABLE3}, the results in the end of
the simulation for $N=36$ are presented, whereas Table IV lists the
results for energy and angular momentum radiated by gravitational
waves.  Here, the values of $\Delta E$ and $\Delta J$ depend on the
radii of the wave extraction by 1--3\% for $N=36$. In addition, these
systematically increase with improving the grid resolution. Thus, 
we infer these quantities by an extrapolation of the data for 
$N=30$ and 36, which carried out assuming the third-order convergence.
(We note that the Einstein and hydrodynamic equations are solved 
in the fourth and third order schemes.)

Table \ref{TABLE3} shows that $M_{\rm BH,f}$ agrees with $C_e/4\pi$
within $\sim 0.5\%$ error, but $M_{\rm BH,f}$ is systematically larger
than $C_e/4\pi$. The likely reason that $\Delta E$, which is used in
computing $M_{\rm BH,f}$, is slightly underestimated for $N=36$
because of numerical dissipation of gravitational wave amplitude:
Indeed, numerical results for the energy and angular momentum radiated
by gravitational waves increase with improving the grid resolution as
mentioned above, and for the extrapolated results shown in Table IV 
another conservation relations such as 
\beqn
{C_e \over 4\pi} + M_{r> r_{\rm AH}} + \Delta E = M_0
\eeqn
holds in a good manner within the numerical error of $\sim 0.1$--0.2\%.

As described in Ref.~\cite{ST08}, there are at least three methods for
approximately estimating the final BH spin. In this paper, we use the
following methods.  In the first method, we approximately
estimate the mass and spin of the BH by conservation laws. Namely, we
determine them by subtracting total energy and angular momentum
dissipated by gravitational waves, and rest mass and angular momentum
of disks surrounding the BH from the initial ADM mass and angular
momentum, respectively. As shown in Eq. (\ref{mass_bhf}), the BH mass
is estimated to give $M_{\rm BH,f}$. In the same manner, angular
momentum of the BH may be estimated by
\beqn
J_{\rm BH,f} \equiv J_0 - J_{r> r_{\rm AH}} - \Delta J, \label{ang_bhf}
\eeqn
where $\Delta J$ is total angular momentum radiated 
by gravitational waves and $J_{r> r_{\rm AH}}$ is angular momentum
of the material located outside the apparent horizon, defined by
\beqn
J_{r> r_{\rm AH}} \equiv \int_{r> r_{\rm AH}} \rho_* h u_{\varphi} d^3x. 
\label{j_disk}
\eeqn
Here, $u_{\varphi}=(x-x_{\rm P})u_y+(y-y_{\rm P})u_x$ and 
$(x_{\rm P}, y_{\rm P})$ denote the position of the puncture. 
$J_{r> r_{\rm AH}}$ exactly gives the angular momentum of the material in
the axisymmetric and stationary spacetime. In the late phase of the
merger, the spacetime relaxes to a quasistationary and nearly
axisymmetric state.  Thus, we may expect that $J_{r> r_{\rm AH}}$
will provide an approximate magnitude of the angular momentum of disks.
From $J_{\rm BH,f}$ and $M_{\rm BH,f}$, we define a nondimensional
spin parameter by $a_{\rm f1} \equiv J_{\rm BH,f}/M_{\rm BH,f}^2$ (see
Table III). 

In the second method, the spin is determined from the following geometric
quantities of the apparent horizon; $C_e/4\pi$ and $M_{\rm irr}$.  For
Kerr BHs of spin $a$, the following relation holds,
\beqn
M_{\rm irr}={C_e \over 4\sqrt{2}\pi}
\Bigl(1+\sqrt{1-a^2}\Bigr)^{1/2}, \label{relK2}
\eeqn
and hence, $a$ is determined from $C_e$ and $M_{\rm irr}$.  We
estimate the spin assuming that Eq. (\ref{relK2}) holds even for the BH
surrounded by the disk \cite{bhdisk07}.  The BH spin
determined by this method is referred to as $a_{\rm f2}$.

In the third method, $C_p/C_e$ is used. For Kerr BHs, it is 
calculated to give 
\beqn
{C_p \over C_e}={\sqrt{2  \hat r_+} \over \pi}
E(a^2/2 \hat r_+), \label{eq:cpce}
\eeqn
where $\hat r_+=1+\sqrt{1-a^2}$ and $E(z)$ is an elliptic
integral defined by
\beq
E(z)=\int^{\pi/2}_0 \sqrt{1-z\sin^2\theta} d\theta.
\eeq
Thus, assuming that the same relation holds even for the BH surrounded by 
the disk, we may estimate a BH spin from $C_p/C_e$. We refer to this 
spin as $a_{\rm f3}$. 

The values of $a_{\rm f1}$--$a_{\rm f3}$ for $N=36$ are listed in
Table III. We find that three values agree within a few \% for
$Q=3$--5. The values of the spin depend weakly on the compactness of
the NSs, and hence, we conclude that the spin parameter of the formed
BH is $\approx 0.56 \pm 0.01$, $0.48 \pm 0.01$, and $0.42 \pm 0.01$
for $Q=3$, 4, and 5, respectively.  For the smaller values of $Q$, the
final BH spin is larger.  The reason for this is that for the larger
mass ratio, the total angular momentum of the system $J$ at a given
value of $m_0\Omega$ in the inspiral orbit is approximately
proportional to $m_0^2 Q/(1+Q)^2$ (see also the initial condition in
Table I). Thus, the binaries of smaller values of $Q$ should form a BH
of higher spin.

For the case that $Q \leq 2$ and the disk mass is $\agt 0.01M_*$,
$a_{\rm f1}$ does not agree well with $a_{\rm f2}$ and $a_{\rm f3}$
with the error size $\sim 0.05$. Because $a_{\rm f2}$ and $a_{\rm f3}$
agree well, the error in $a_{\rm f1}$ seems to be much larger than
those of $a_{\rm f2}$ and $a_{\rm f3}$. Then, the possible error
sources are (i) underestimation of angular momentum that the disks possess
and/or (ii) underestimation of angular momentum dissipated by
gravitational waves. The possibility (ii) is not very likely, because
for $Q \geq 3$ and for models M20.160 and M20.178, $a_{\rm
  f1}$--$a_{\rm f3}$ agree in a much better manner, indicating that
gravitational waves are computed with a good accuracy.  A possible
reason that the angular momentum of the disk is underestimated is that
many of the disk materials are located not in the finest grid domain
in the AMR grid but in the second--fourth finest domains in which the
grid resolution may not be high enough, and hence, the angular momentum
is spuriously dissipated. This reason is also inferred from
Fig.~\ref{FIG7} (d), which shows that the disk mass surrounding BH
depends on the grid resolution.

\begin{table*}[t]
\caption{Rest mass of material located outside apparent horizon ($M_{r
> r_{\rm AH}}$), BH mass estimated by energy-conservation law ($M_{\rm
BH,f}$), BH mass estimated from equatorial circumferential radius
($C_e/4\pi$), irreducible mass of the BH ($M_{\rm irr}$; square root
of area of apparent horizon in units of $16\pi$), ratio of polar
circumferential radius ($C_p$) to equatorial one ($C_e$) of apparent
horizon (i.e., $C_p/C_e$), and estimated spin parameters of the final
state of the BH. $a_{\rm f1}$, $a_{\rm f2}$, and $a_{\rm f3}$ are
computed from BH mass and angular momentum estimated by conservation
laws, from $M_{\rm irr}$ and $C_e$ of apparent horizon, and from
$C_p/C_e$, respectively. All the values presented here are measured
for the state obtained at the end of the simulations for $N=36$.  Note
that the parameters for the BH still vary with time at the end of
the simulation for models M15.145, M20.145, and M20.145N because of
gradual mass accretion. The error for $C_e$ and
$C_p/C_e$ is $\alt 0.1\%$, whereas that for $a_{\rm f1}$, $a_{\rm
f2}$, and $a_{\rm f3}$ is $\alt 0.01$ except for the case that the disk is
formed for which the error of $a_{\rm f1}$ would be $\sim 0.05$. 
The values for the final state of the BH depend very weakly on the 
grid resolution as far as $N \geq 30$, but the mass of 
disk which presents only for $Q \leq 2$ systematically increases 
with $N$. The present results should be regarded as the lower-bound 
for the disk mass. 
\label{TABLE3}}
\begin{tabular}{ccccccccc} \hline
Model & $M_{r > r_{\rm AH}}/M_*$ & $M_{\rm BH,f}/M_0$ & $C_e/4\pi
M_0$ & $M_{\rm irr}/M_0$ & $C_p/C_e$ & $a_{\rm f1}$ & 
$a_{\rm f2}$ & $a_{\rm f3}$ \\ \hline
M15.145  & 0.023 
& 0.983 & 0.981 & 0.895 & 0.872 & ~0.801 & ~0.747 & ~0.750 \\ \hline
M20.145  & 0.010 
& 0.988 & 0.984 & 0.915 & 0.898 & ~0.717 & ~0.684 & ~0.682 \\ \hline
M20.145N & 0.011 
& 0.988 & 0.983 & 0.916 & 0.900 & ~0.721 & ~0.677 & ~0.676 \\ \hline 
M20.160 & $6 \times 10^{-4}$ 
& 0.988 & 0.987 & 0.919 & 0.899& ~0.694 & ~0.679 & ~0.680 \\ \hline
M20.178 & $< 10^{-5}$ 
& 0.983 & 0.983 & 0.917 & 0.904& ~0.676 & ~0.672 & ~0.666 \\ \hline
M30.145  & $<10^{-5}$ 
& 0.989 & 0.985 & 0.942 & 0.934& ~0.566 & ~0.559 & ~0.564 \\ \hline 
M30.160  & $<10^{-5}$
& 0.985 & 0.983 & 0.940 & 0.935& ~0.547 & ~0.560 & ~0.562 \\ \hline
M30.178  & $<10^{-5}$ 
& 0.982 & 0.981 & 0.940 & 0.937 & ~0.551 & ~0.550 & ~0.552 \\ \hline
M40.145  & $<10^{-5}$ 
& 0.988 & 0.986 & 0.955 & 0.953& ~0.485 & ~0.482 & ~0.474 \\ \hline 
M50.145  & $<10^{-5}$ 
& 0.989 & 0.986 & 0.963 & 0.964& ~0.408 & ~0.419 & ~0.425 \\ \hline 
\end{tabular}
\end{table*}

\begin{table*}[t]
\caption{Several outputs of gravitational waves.  Total radiated
  energy ($\Delta E$) and angular momentum ($\Delta J$) in units of
  initial ADM mass ($M_0$) and initial angular momentum ($J_0$),
  fraction of radiated energy for $(l, |m|)=(2,2)$, (3,3), (4,4), and
  (2,1) modes, frequency of the fundamental quasinormal mode, kick
  velocity, and type of the gravitational waveform. The radiated
  energy for each mode is shown in unit of $M_0$ by $\%$. The origin
  of the error bar is primarily the numerical error associated with
  the finite grid resolution, and in part, the finite extraction
  radii of gravitational waves. Note that gravitational waves are 
  extracted for several coordinate radii of $50$--$100M_0$.
\label{TABLE4}}
\begin{tabular}{cccccccccc} \hline
Model & $\Delta E/M_0$ (\%) & $\Delta J/J_0$ (\%) 
& ~~~~~~(2,2)~~~~~~ & ~~~~~~(3,3)~~~~~~ & ~~~~~~(4,4)~~~~~~ & ~~~~~~(2,1)~~~~~~
& $f_{\rm QNM} M_{\rm BH}$ & $V_{\rm kick}$ (km) & Type \\ \hline
M15.145  & $0.68 \pm 0.02$ & $14 \pm 1$ &$0.66 \pm 0.02$ & $0.006\pm 0.002$
&$0.003 \pm 0.001$&$\alt 0.01$ & --- & $<5$ & I \\ \hline
M20.145  & $0.87 \pm 0.02$ & $17.4 \pm 0.3$ &$0.85 \pm 0.01$&$0.017\pm 0.002$
&$0.005 \pm 0.001$ &$0.003 \pm 0.001$ & ---  & $21 \pm 5$ & I \\ \hline
M20.145N & $0.78 \pm 0.02$ & $15.0 \pm 0.2$ &$0.76 \pm 0.01$&$0.014\pm 0.001$ 
& $0.005 \pm 0.001$ &$0.002 \pm 0.001$ & ---  & $11 \pm 2$ & I \\ \hline 
M20.160  & $1.22 \pm 0.02$ & $22 \pm 1$ & $1.19 \pm 0.02$ & $0.025 \pm 0.005$ 
&$0.006 \pm 0.002$ & $\alt 0.01$ & --- & $62 \pm 15$ & II \\ \hline
M20.178  & $1.7 \pm 0.1$ & $25 \pm 1$ & $1.6 \pm 0.1$ & $0.05 \pm 0.01$ 
&$0.01 \pm 0.005$ & $0.03 \pm 0.01$ & 0.087 & $126 \pm 20$ & II \\ \hline
M30.145  & $1.3 \pm 0.1$ & $22.0 \pm 0.4$ & $1.2 \pm 0.1$ & $0.08 \pm 0.01$ 
&$0.014 \pm 0.002$ & $0.03 \pm 0.02$ & 0.081 &$98 \pm 9$  & II \\ \hline 
M30.160  & $1.6 \pm 0.1$ & $26 \pm 1$ & $1.4 \pm 0.1$ & $0.09 \pm 0.02$ 
&$0.020 \pm 0.003$& $0.05 \pm 0.02$ & 0.080 &$153 \pm 16$  & III\\ \hline
M30.178  & $1.8 \pm 0.1$ & $26 \pm 1$ & $1.7 \pm 0.1$ & $0.12 \pm 0.01$ 
&$0.02 \pm 0.01$& $0.04 \pm 0.03$ & 0.080 &$137 \pm 43$ & III \\ \hline
M40.145  & $1.3 \pm 0.1$ & $23.5 \pm 0.5$ & $1.1 \pm 0.1$ & $0.13 \pm 0.01$ 
& $0.025 \pm 0.005$ & $0.06 \pm 0.02$ & 0.077 &$136 \pm 12$ & III \\ \hline 
M50.145  & $1.11 \pm 0.05$ & $24 \pm 1$ & $0.85 \pm 0.02$ & $0.13 \pm 0.01$ 
& $0.04 \pm 0.01$ & $0.09 \pm 0.01$ & 0.074 &$137 \pm 6$ & III \\ \hline 
\end{tabular}
\end{table*}

\subsection{Gravitational waves}\label{sec:gw}

\subsubsection{Comparison with Taylor T4 waveform}

\begin{figure*}[t]
\vspace{-8mm}
\begin{center}
\epsfxsize=3.5in
\leavevmode
\epsffile{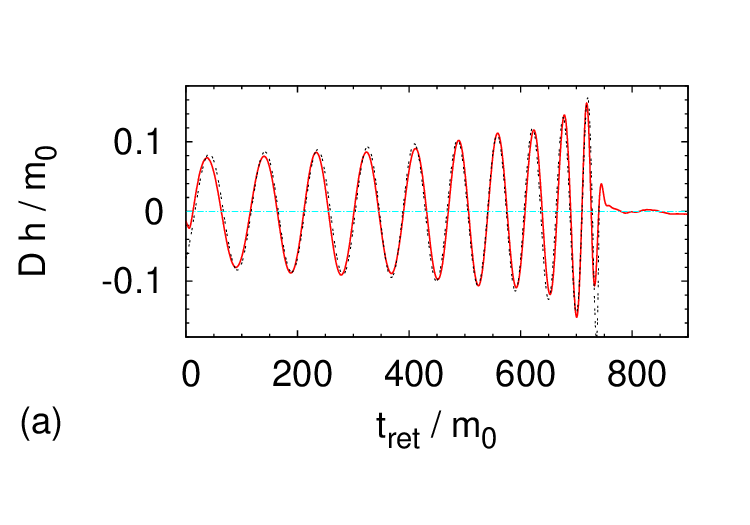}
\epsfxsize=3.5in
\leavevmode
~\epsffile{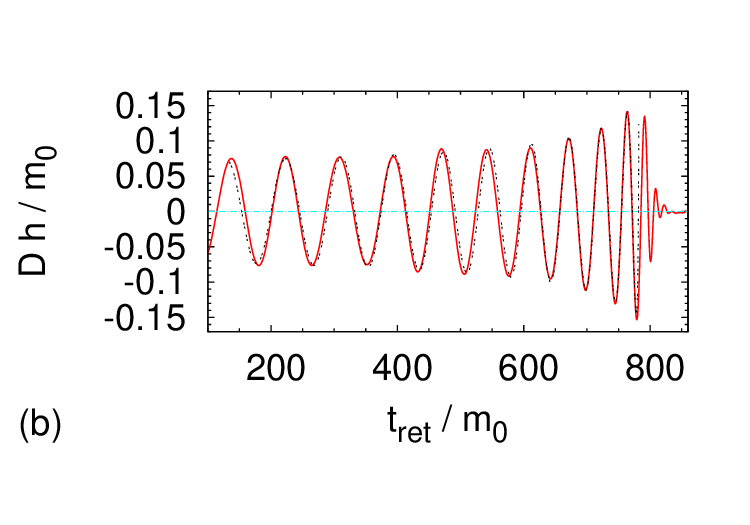} \\
\vspace{-9mm}
\epsfxsize=3.5in
\leavevmode
\epsffile{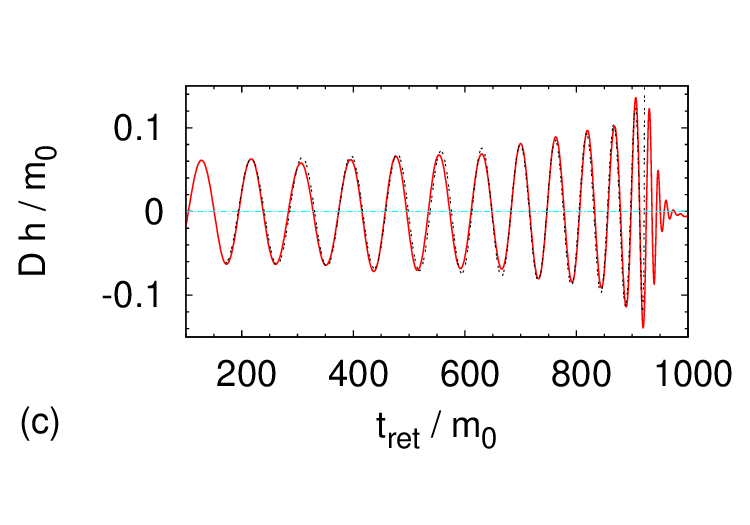}
\epsfxsize=3.5in
\leavevmode
~\epsffile{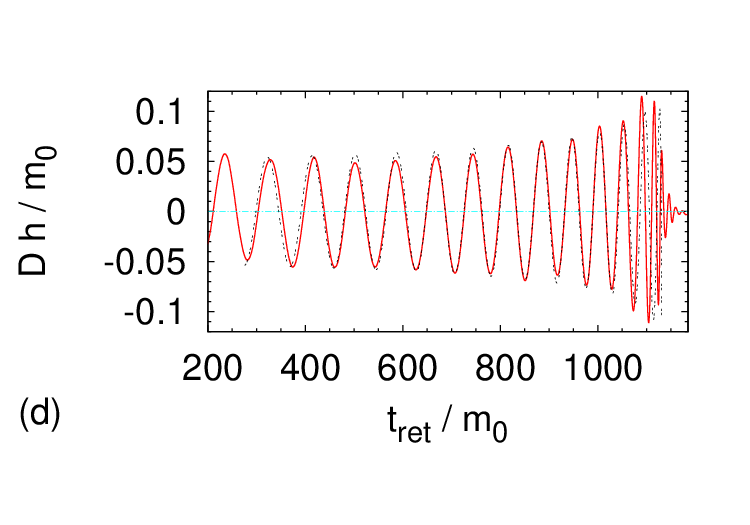} \\
\vspace{-9mm}
\epsfxsize=3.5in
\leavevmode
\epsffile{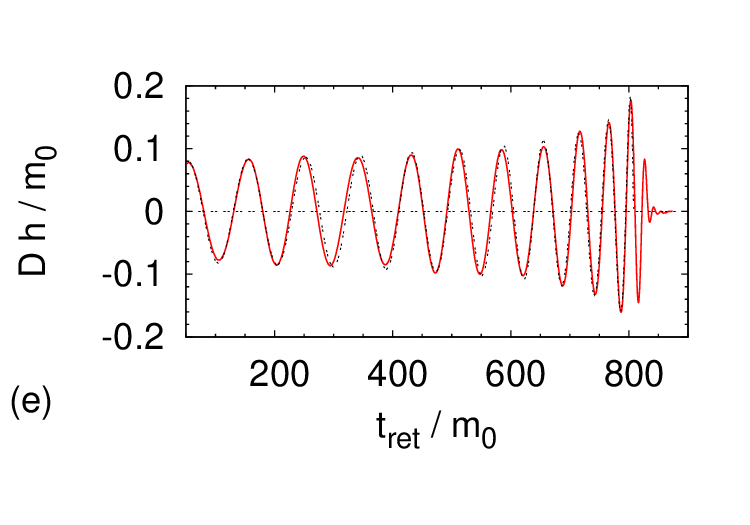}
\epsfxsize=3.5in
\leavevmode
~\epsffile{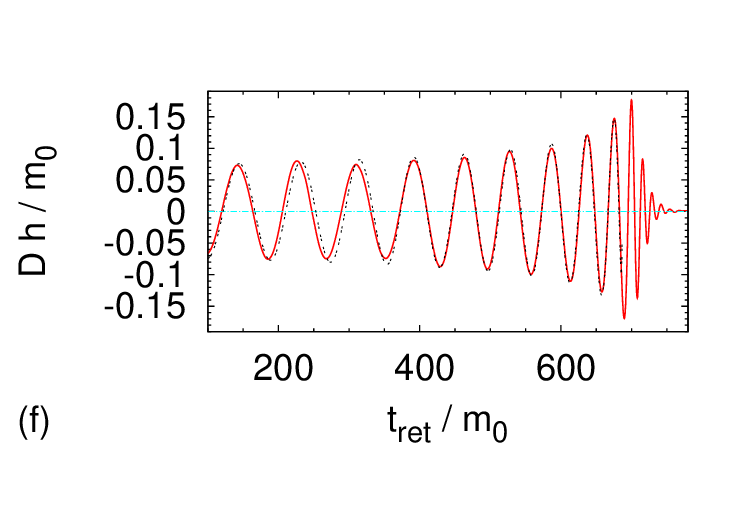} 
\end{center}
\vspace{-12mm}
\caption{Gravitational waveforms observed along the $z$ axis (solid
curve) for models (a) M20.145, (b) M30.145, (c) M40.145, (d) M50.145,
(e) M20.160, and (f) M30.178.  $t_{\rm ret}$ denotes the retarded time
[see Eq. (\ref{ret}) for definition] and $m_0$ is the total mass
defined by $M_{\rm BH}+M_{\rm NS}$.  The amplitude at a hypothetical
distance $D$ can be found from Eq. (\ref{hamp}). The dot-dotted curves
denote the waveforms derived by the Taylor-T4 formula.
\label{FIG9}}
\end{figure*}

\begin{figure*}[t]
\vspace{-8mm}
\begin{center}
\epsfxsize=3.3in
\leavevmode
\epsffile{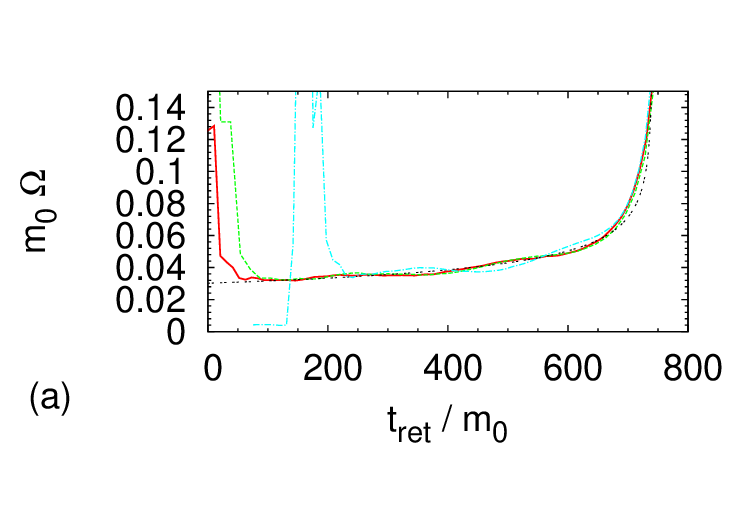}
\epsfxsize=3.3in
\leavevmode
~~\epsffile{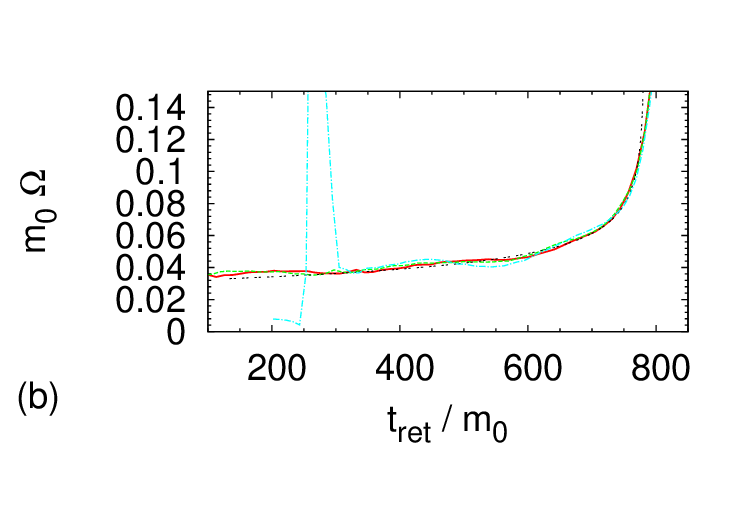} \\
\vspace{-1.2cm}
\epsfxsize=3.3in
\leavevmode
\epsffile{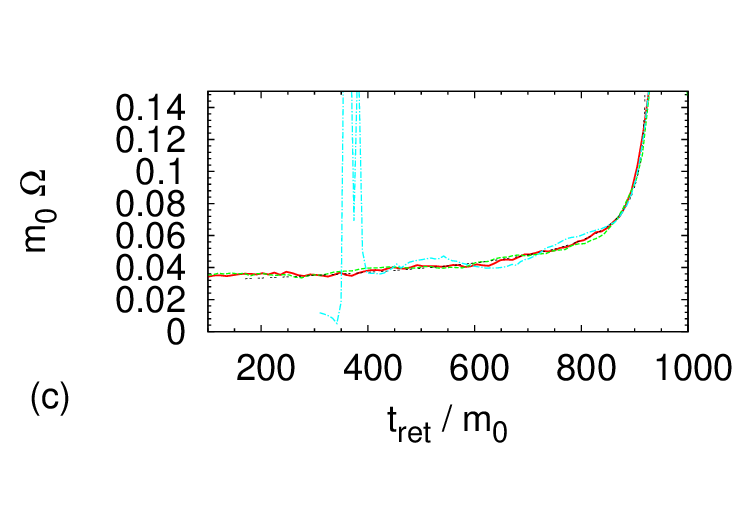}
\epsfxsize=3.3in
\leavevmode
~~\epsffile{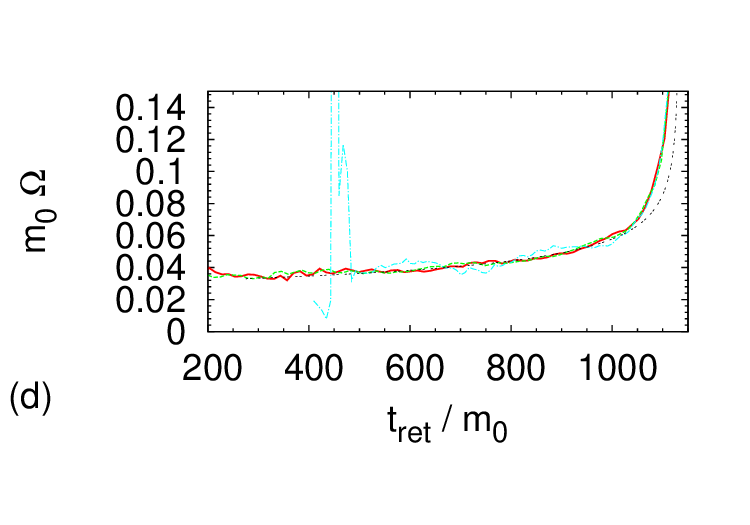}
\end{center}
\vspace{-13mm}
\caption{Orbital angular velocity computed from $\Psi_4$ 
as a function of the retarded time for models (a) M20.145, (b)
M30.145, (c) M40.145, and (d) M50.145.  $t_{\rm ret}$ denotes the
retarded time and $m_0$ is the total mass.  For all the panels, the
results for two different grid resolutions with $N=30$ (dashed curve)
and 36 (solid curve) and the results for M20.145b--M50.145b
(dot-dashed curves) are shown together. The dot-dotted curve denotes
the result derived by the Taylor-T4 formula. 
\label{FIG10}}
\end{figure*}

Figure \ref{FIG9} (a)--(f) plot gravitational waveforms ($+$ mode) 
observed along the $z$ axis as a function of the retarded time for models
M20.145--M50.145, M20.160, and M30.178.  ($h$ denotes a gravitational
wave amplitude.) The retarded time is approximately defined by
\beqn
t_{\rm ret} \equiv t - D - 2M_0 \ln (D/M_0), \label{ret}
\eeqn
where $D$ is a distance between the source and
an observer.
The gravitational waveforms shown here are obtained by performing the 
time integration of the Newman-Penrose quantity of $l=|m|=2$ mode. 
From the values of $D h/m_0$ shown in Fig.~\ref{FIG9}, 
the amplitude of gravitational waves at a distance $D$ is evaluated by 
\beqn
&&h_{\rm gw} \approx 2.4 \times 10^{-22} 
\biggl( {D h/m_0 \over 0.1}\biggr) \nonumber \\
&& ~~~~~~~~~~~~~~~~~~~\times 
\biggl({100~{\rm Mpc} \over D}\biggr)
\biggl({m_0 \over 5 M_{\odot}}\biggr). \label{hamp}
\eeqn

To validate the numerical waveforms presented here, we first compare
the waveforms in the inspiral phase with those derived by the
so-called Taylor-T4 formula for two point masses in quasicircular
orbits. In the Taylor-T4 formula, one calculates evolution of the
angular velocity, $\Omega$, of the quasicircular orbits due to the
gravitational radiation reaction up to the 3.5PN level beyond the
quadrupole formula: The circular orbits at a given value of $\Omega$
are determined by the 3PN equations of motion neglecting gravitational
radiation reaction, and then, one considers an adiabatic evolution of
$\Omega$ using the 3.5PN formula for gravitational radiation reaction
(see, e.g., Refs.~\cite{BHBH12,BHBH} for a detailed description of
various formulas based on the PN theory). Recent high-accuracy
simulations for equal-mass (nonspinning or corotating) BH-BH binaries
have proven that the Taylor-T4 formula provides their orbital
evolution and gravitational waveforms with a high accuracy at least up
to about one orbit before the onset of the merger. Assuming that this
holds for unequal-mass binaries, we calibrate our numerical results by
comparing them with the results by the Taylor-T4 formula.  (Indeed,
our numerical results indicate that the Taylor-T4 formula provides a
good approximate solution for $\Omega(t)$ even for the nonequal-mass
binaries as shown below.) In the present work, the 3PN formula
\cite{Kidder} is employed for calculating the amplitude of
gravitational waves in the Taylor-T4 formula.

More specifically, the comparison of the numerical waveforms and
semianalytic ones derived by the Taylor-T4 formula is carried out in
the following manner.  First, we derive the orbital angular velocity
as a function of time for a numerical result from $\Psi_4$ by
\cite{BHBH12}
\beqn
\Omega(t)
={1 \over 2} {|\Psi_4(l=m=2)| \over \displaystyle \Big|\int dt
\Psi_4(l=m=2)\Big|},
\label{gwangv}
\eeqn 
where $\Psi_4(l=m=2)$ is the $l=m=2$ mode of $\Psi_4$. Then, we
compare the numerical result for $\Omega(t)$ with the semianalytic one
derived by the Taylor-T4 formula (see Fig.~\ref{FIG10}). When
comparing two results, we have a degree of freedom for the time 
translation.  Thus, first of all, by shifting the time axis of the 
Taylor-T4's result, we align the origin of the time.  As shown in 
Fig.~\ref{FIG10}, we can always shift the time axis appropriately to 
align two results for $\Omega(t)$. 

After the appropriate time translation, we compare the gravitational
waveforms obtained by the numerical simulation and by the Taylor-T4
formula. Then, we still have a degree of freedom for choosing the
wave phase. Thus, we iteratively change the phase of the Taylor-T4's
waveform until a good matching of two waveforms is achieved.

\begin{figure*}[thb]
\vspace{-8mm}
\begin{center}
\epsfxsize=3.3in
\leavevmode
\epsffile{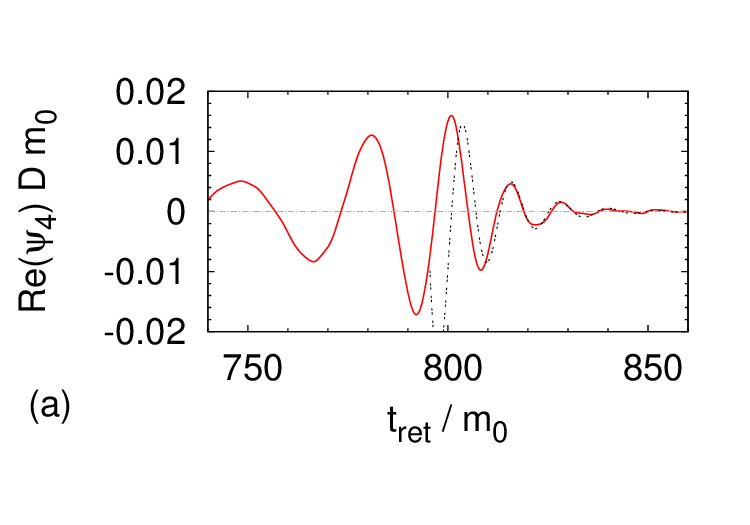}
\epsfxsize=3.3in
\leavevmode
~~~~~~~\epsffile{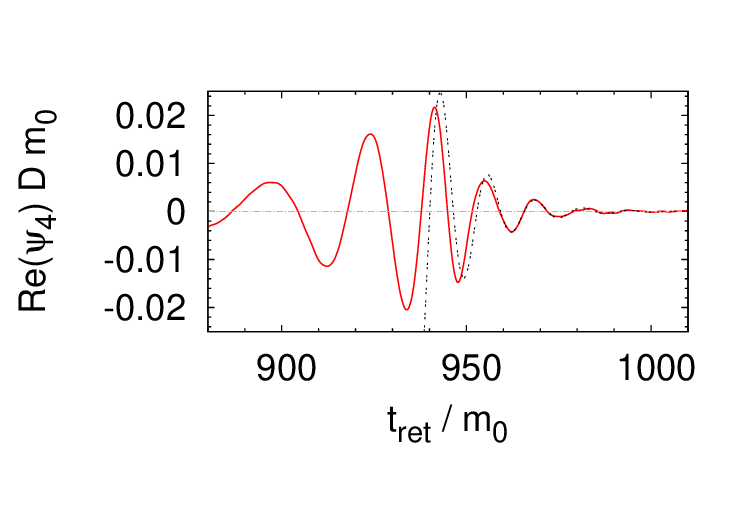} \\
\vspace{-1.2cm}
\epsfxsize=3.3in
\leavevmode
\epsffile{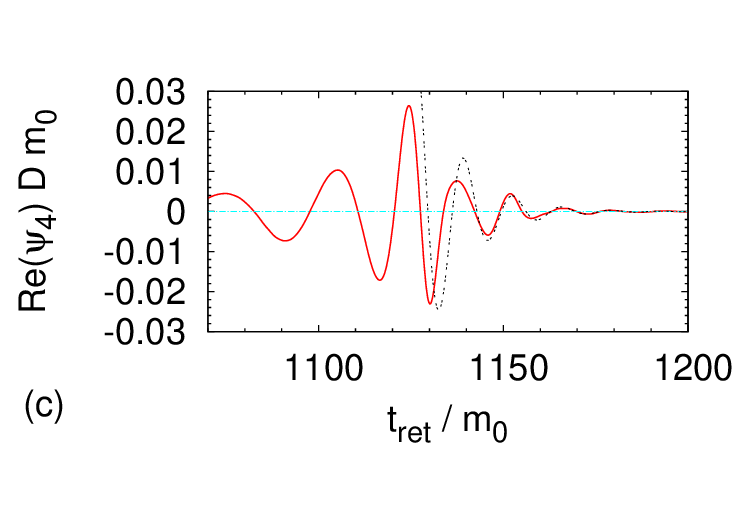}
\epsfxsize=3.3in
\leavevmode
~~~~~~~\epsffile{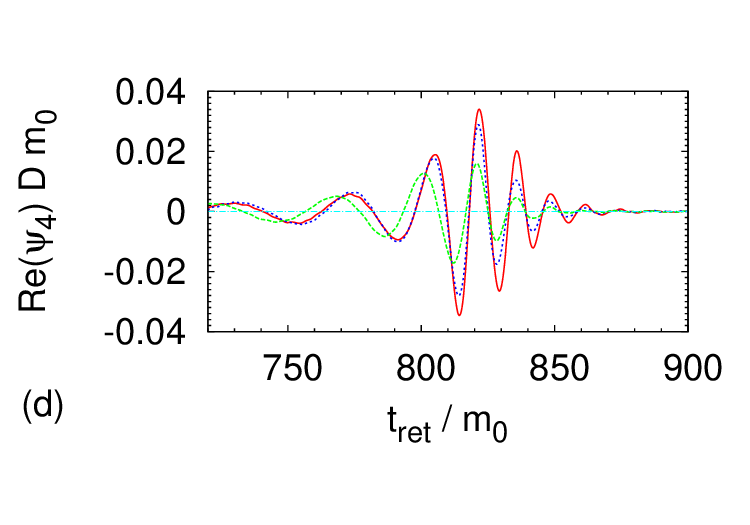}
\end{center}
\vspace{-13mm}
\caption{Gravitational waves (the real part of $\Psi_4$) emitted in
the merger and ringdown phases for models (a)
M30.145, (b) M40.145, (c) M50.145, and (d) M30.145 (dashed curve),
M30.160 (dotted curve), and M30.178 (solid curve). For (a)--(c), a
fitting waveform given by Eq. (\ref{QNM}) is plotted together by the
dot-dotted curve. 
\label{FIG11}}
\end{figure*}

In Fig.~\ref{FIG9}, the dot-dotted curve denote the resulting
semianalytic waveform derived by the Taylor-T4 formula. It is found
that the numerical waveforms agree with the Taylor-T4's results with a
good accuracy, except for the early stage of the simulations (i.e.,
for the first a few wave cycles), during which the eccentricity of the
binaries is not small. In particular, for the last several wave cycles
(except for the orbit just before the merger), the numerical wave
phases agree with those derived by the Taylor-T4 formula with in $\sim
3\%$ error.  This indicates that the binaries computed in the present
simulation are indeed in an approximately quasicircular orbit of small
eccentricity at least for the last several inspiral orbits. Also, this
indicates that the Taylor-T4 formula provides a good approximate
solution for $\Omega(t)$ even for the nonequal-mass binaries, because 
the agreement systematically holds irrespective of the mass 
ratio. 

Figure \ref{FIG10} also shows good agreement between the numerical and
Taylor-T4's results for $\Omega(t)$. The numerical results for two
grid resolutions [$N=36$ (solid curve) and 30 (dashed curve)] are
shown for illustrating that a convergence is approximately
achieved. In addition, the results for models M20.145b--M50.145b are
plotted for comparison.  This figure shows that the evolution of
$\Omega(t)$ for models M20.145--M50.145 agrees well with those
predicted by the Taylor-T4 formula for a long time duration
irrespective of the mass ratio.  Modulation in the evolution of
$\Omega$ is $\Delta \Omega/\Omega \alt 6\%$ for the last a few
inspiral phase (irrespective of $N=30$ or 36), and thus, the
eccentricity, which is approximately estimated by $2\Delta
\Omega/3\Omega$, is $\alt 4\%$. In the last several orbits, the
eccentricity appears to be at most $\sim 1\%$.


Comparing the results of M20.145--M50.145 with those of
M20.145b--M50.145b, we find that the modulation is by a factor of 
$\agt 2$ larger with the initial condition computed in the
$\beta^{\varphi}$ condition. This unfavored behavior is more 
significant for the larger values of $Q$. This demonstrates the
advantage for using the initial condition computed in the 3PN-J
condition. 

\subsubsection{Classification of waveforms}

Figure \ref{FIG9} shows that gravitational waveforms in the merger and
ringdown phases depend sensitively on the mass ratio and compactness
of the NS.  Comparison of the waveforms for models M20.145 and M20.160
[see Fig.~\ref{FIG9} (a) and (e)], for which masses of the BH and NS
are identical each other, illustrates a strong dependence of the
merger waveforms on the NS compactness. The wave amplitude for model
M20.145 decreases suddenly in the middle of the inspiral phase due to
tidal disruption. By contrast, the wave amplitude for model M20.160
does not decrease as quickly as that for model M20.145 because the
tidal disruption does not occur far outside the ISCO.

Gravitational waveforms in the merger and ringdown phases for models
M30.145 and M30.178 [see Fig.~\ref{FIG9} (b) and (f), and
Fig.~\ref{FIG11}(d)] are also distinguishable because the amplitude in
these phases is much larger for model M30.178.  This is due to the
fact that the NS for model M30.178 is not strongly affected by the
tidal force of the companion BH even at the ISCO.  By contrast, the
tidal effects play an important role for deformation of the NS in
close orbits for model M30.145. Because the NS is disrupted near the
ISCO for this model, the wave amplitude in the merger and ringdown
phases is significantly suppressed.  These results illustrate that
gravitational waves emitted in the merger and ringdown phases have a
potential information about the compactness of the NS (see also 
Sec. \ref{spectrum}). 

As these comparisons clarify, there are three types of gravitational
waveforms.  For the case that the NS is tidally disrupted during the
inspiral phase (e.g., for model M20.145), the wave amplitude quickly
decreases at the tidal disruption, as the waveforms associated with
the inspiral motion are suddenly shut off. Namely, the waveform is
composed only of the inspiral waveform and subsequent sudden shut-off,
and the merger and ringdown waveforms are essentially absent. We
refer to this type as the type I. 

Even in this case, most of the material of the tidally disrupted NS
falls into the companion BH (see Fig.~\ref{FIG7}). During the tidal
disruption and subsequent infalling into the BH, an orbital motion of
the disrupted material and an oscillation of the BH may excite merger
and ringdown gravitational waves (here ``ringdown gravitational
waves'' imply gravitational waves associated with quasinormal modes of
the BH).  However, such waveforms are not seen. The likely reason for
the absence of the merger waveform is that the NS is significantly
elongated in a short time scale and its density quickly decreases,
suppressing an efficient excitation of gravitational waves. The reason
for the absence of the ringdown waveform is that the material of the
NS, which falls into the BH, does not have a compact configuration but
have an elongated low-density configuration. In the case that such
low-density diffuse matter incoherently falls into the BH, the
excitation of the quasinormal modes is significantly suppressed due to
the phase cancellation effect \cite{SaNa}.

For models M30.145 and M20.160, mass shedding occurs before the
binaries reach the ISCO. However, the sudden shut-off of the inspiral
waveforms is not seen because the tidal disruption and the subsequent
spreading of the material do not occur during the inspiral
phase. Rather, most part of the elongated NS falls into the BH before
the tidal disruption is completed. In this case, the ringdown waveform
is seen but the amplitude is low because the quasinormal mode is not
excited efficiently. By contrast, just before the ringdown
gravitational waves are emitted, gravitational waves are significantly
excited by a matter motion around the BH.  Thus, the merger
gravitational waves are present. Namely, in these cases, the
gravitational waveforms are composed primarily of the inspiral and
merger ones. We refer to this type as the type II.

For models M40.145, M50.145, and M30.178, tidal effects to the NS do
not play an important role. In this case, the gravitational waveform
is composed of the inspiral, merger, and ringdown waveforms, as in the
merger of BH-BH binaries (e.g., \cite{BHBH12,BHBH}). We refer to this
type as the type III.

\subsubsection{Ringdown waveforms}

For models M30.145, M40.145, M50.145, M30.160, M20.178, and M30.178,
ringdown gravitational waves associated with quasinormal modes of the
formed BH are excited in the final phase. These gravitational waves
are approximately described by
\beqn
A e^{-t/t_d} \sin(2\pi f_{\rm QNM} t + \delta), \label{QNM}
\eeqn
where $A$ and $\delta$ are constants, and $f_{\rm QNM}$ and $t_d$ are
the frequency and damping time scale of the fundamental quasinormal
mode.  A perturbation study predicts the frequency and damping time
scale for a BH of mass $M_{\rm BH}$ and spin $a$ as \cite{leaver}
\beqn
&& f_{\rm QNM} M_{\rm BH} \approx 0.16
[ 1-0.63(1-a)^{0.3}], \label{eqbh} \\
&& t_d \approx {2(1-a)^{-0.45} \over \pi f_{\rm QNM}}. 
\label{QNMf}
\eeqn
For models M30.145, M40.145, and M50.145, the BH spin is $\approx
0.56$, 0.48, and 0.42 as shown in Sec. \ref{BHspin}. (For models
M30.160 and M30.178, the spin agrees approximately with that for model
M30.145.)  Thus, for each of these models, $f_{\rm QNM} M_{\rm BH}
\approx 0.081$, 0.077, and 0.074 ($f_{\rm QNM} M_{0} \approx 0.082$,
0.078, and 0.075), and $\pi t_d f_{\rm QNM} \approx 2.9$, 2.7, and
2.6, respectively. 

Figure~\ref{FIG11} (a)--(c) compares the numerical waveforms in
the ringdown phase with the hypothetical analytic waveforms for models
M30.145--M50.145. This shows that the numerical waveforms are fitted
by the analytic one (\ref{QNM}) fairly well, and that $f_{\rm QNM}$ and $t_d$
computed from the data of the BH geometry agree approximately with
those computed from gravitational waveforms. However, the numerical
waveforms do not agree completely with the hypothetical ones.  This
disagreement is reasonable because for these models, the material of
the NS does not simultaneously fall into the BH at the merger. Thus,
gravitational waves are emitted both by a motion of the material
moving in the vicinity of the BH and by the quasinormal-mode
oscillation of the BH. In addition, the quasinormal modes are not
simultaneously excited because the material does not simultaneously
fall into the BH, and the resulting waveforms may be composed of many
ringdown waveforms, as well as of the waveforms excited by a material
moving around the BH.  Furthermore, the system is not completely in a
vacuum nor in a stationary state, and hence, the numerical waveforms
may not be fitted precisely by the analytic results derived in the 
ideal assumption.

The wavelength of gravitational waves emitted in the merger phase
(around the time when the peak amplitude is reached) is slightly
longer than that of the quasinormal mode (i.e., the frequency is lower
than $f_{\rm QNM}$).  This indicates that these gravitational waves
are not emitted by an oscillation of the BH, but they are likely to be
emitted primarily by a motion of the material which moves in the
vicinity of the BH.  In the merger phase, the amplitude gradually
decreases after the peak is reached. The reason for this behavior is
that the material is elongated by the tidal effect of the BH during
the infalling; i.e., the reason is not the damping associated with the
quasinormal mode oscillation.  The amplitude emitted in the merger
phase is much larger than that emitted in the ringdown phase, although
the characteristic frequencies of these two types of gravitational
waves are not very different.  In Sec. \ref{spectrum}, we find that
the Fourier spectrum has a plateau in a high frequency region $f m_0
\sim 0.06$--0.08 for the case that $Q=3$--5. This plateau is primarily
generated by gravitational waves emitted in the merger phase not by
the quasinormal mode oscillation.

Figure~\ref{FIG11} (d) compares the waveforms emitted in the
merger and ringdown phases for models M30.145, M30.160, and
M30.178. For these models, the masses of the BH and NS are identical
each other, but the compactness of the NSs is different.  This figure
shows that the amplitude is larger for the model with the larger value
of ${\cal C}$.  This is reasonable because more compact NSs are less
subject to the tidal effects by the companion BH, and hence, the
material of the NS falls into the BH in more simultaneous manner, 
resulting in a coherent excitation of gravitational waves.

The difference in the amplitude of gravitational waves emitted in the
merger and ringdown phases is reflected also in the noticeable
difference of energy and angular momentum carried by gravitational
waves.  As shown in Table IV, for example, the total energy radiated
for models M30.160 and M30.178 is by $\sim 35\%$ and $\sim 50\%$
larger than that for model M30.145, respectively.

\subsection{Gravitational wave spectrum}\label{spectrum}

\begin{figure*}[t]
\vspace{-4mm}
\epsfxsize=3.3in
\leavevmode
\epsffile{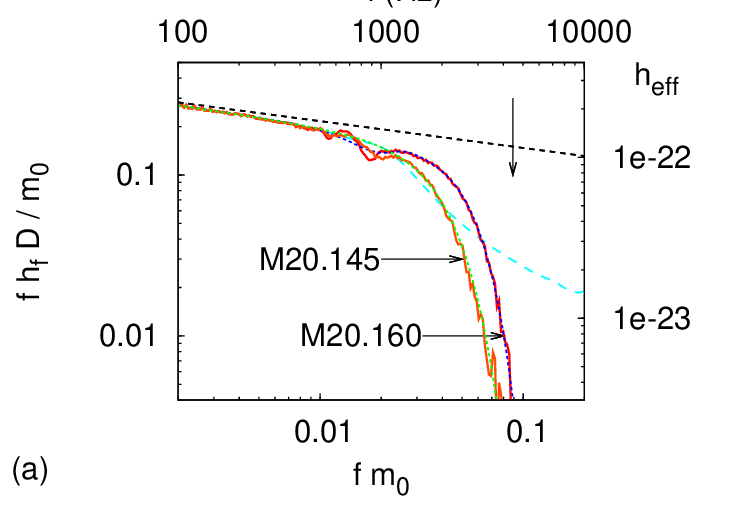}
\epsfxsize=3.3in
\leavevmode
~~~~~\epsffile{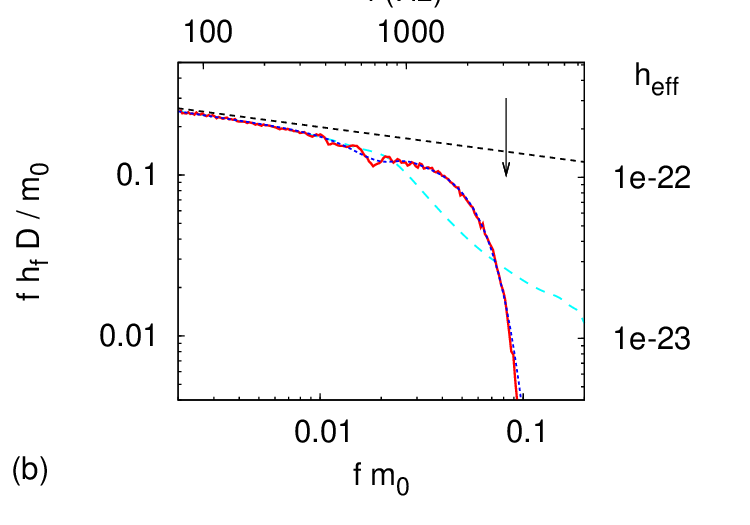}\\
\vspace{3mm}
\epsfxsize=3.3in
\leavevmode
\epsffile{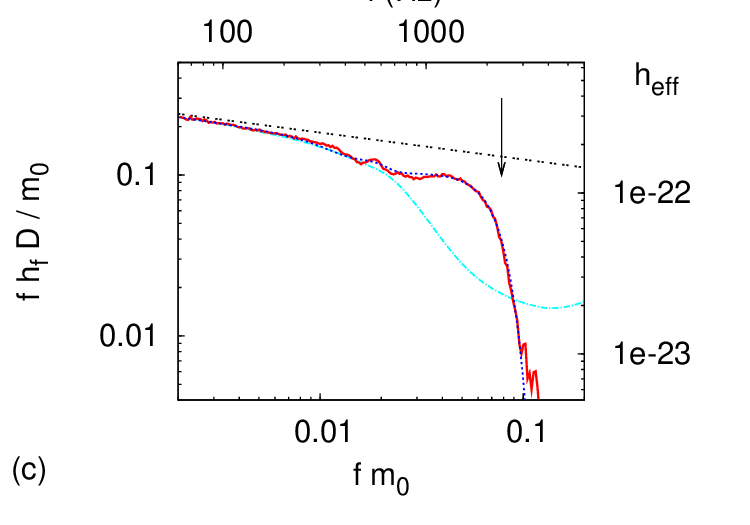}
\epsfxsize=3.3in
\leavevmode
~~~~~\epsffile{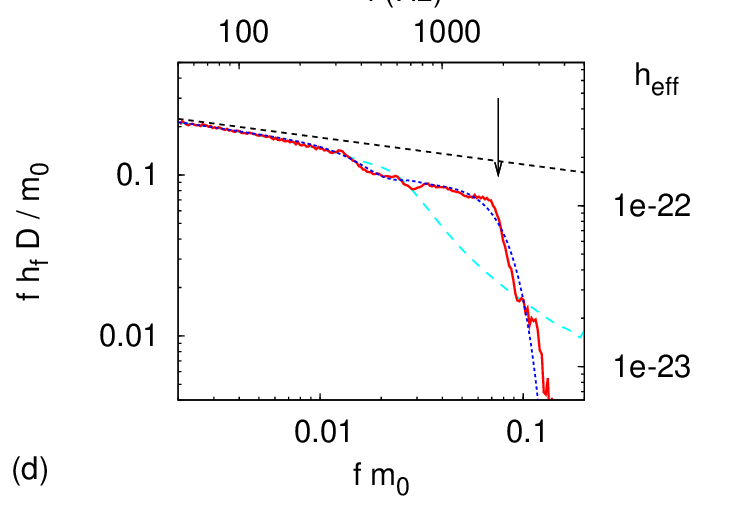}\\
\vspace{3mm}
\epsfxsize=3.3in
\leavevmode
\epsffile{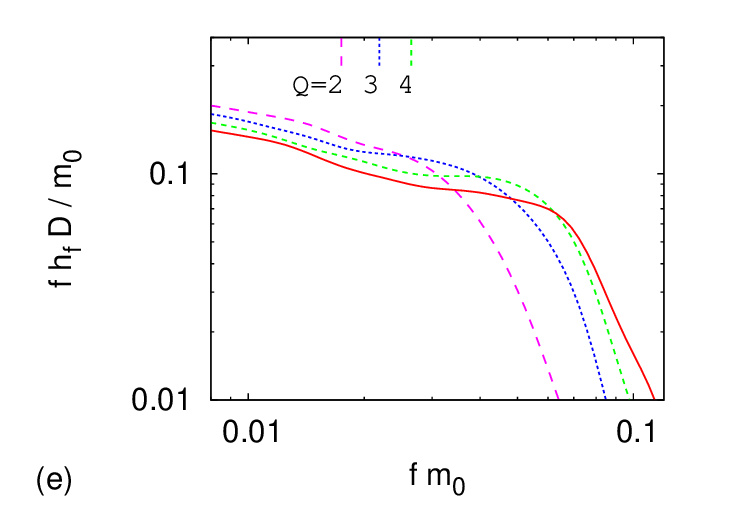}
\epsfxsize=3.3in
\leavevmode
~~~~~\epsffile{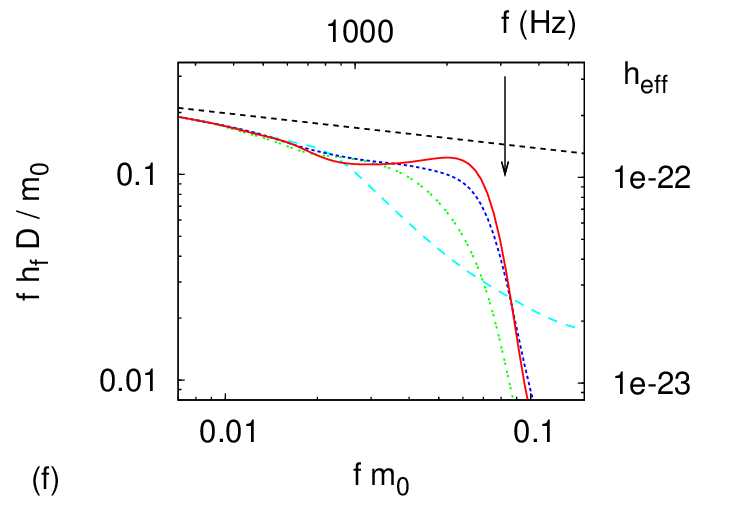}
\vspace{-3mm}
\caption{(a)--(d) The spectrum of gravitational waves $f h(f) D/m_0$
  (solid curves) for models (a) M20.145 and M20.160, (b) M30.145, (c)
  M40.145, and (d) M50.145, respectively. The dashed and long-dashed
  curves denote the relation according to Eq. (\ref{Nspec}) and
  spectra of gravitational waves computed in the Taylor-T4 formula.
  The dotted curves denote the results of fitting. The upper
  horizontal and right vertical axes show the value in a hypothetical
  value of $M_{\rm NS}=1.35M_{\odot}$ and $D=100$ Mpc. The arrow
  indicates the frequency of the fundamental quasinormal mode
  calculated by Eq. (\ref{QNMf}).  (e) The same as (a) but for
  gathering the spectra for models M20.145--M50.145 (long-dashed,
  dotted, dashed, and solid curves).  To clarify the qualitative
  feature of the spectra, a smoothing procedure is applied for the
  numerical data. The three lines in the upper left side denote the
  predicted frequency at which mass shedding of the NS occurs due to
  tidal field of the companion BH for models M20.145--M40.145 ($Q$
  denotes the mass ratio).  (f) The same as (b) but for models M30.145
  (dot-dotted curve), M30.160 (dotted curve), and M30.178 (solid
  curve). A smoothing procedure is also applied for the numerical
  data.
\label{FIG12}}
\end{figure*}

To determine the effective amplitude of gravitational waves for a
given frequency, the Fourier spectrum of gravitational waves of $l=|m|=2$
modes are computed.  In this paper, as the Fourier spectrum, we define
\beqn
h(f) \equiv \sqrt{{|h_+(f)|^2+|h_{\times}(f)|^2 \over 2}},
\eeqn
where $f$ is the frequency, 
$h_+(f)$ and $h_{\times}(f)$ are the Fourier transformation of
the plus and cross modes of gravitational waves observed along the $z$
axis; 
\beqn
&&h_+(f)=\int e^{2\pi i f t} h_+(t)dt,\\
&&h_{\times}(f)=\int e^{2\pi i f t} h_{\times}(t)dt.
\eeqn
Then, the most optimistic effective amplitude of gravitational
waves for a given frequency is defined by $f h(f)$. 

In the numerical simulations, BH-NS binaries of a finite orbital
separation are prepared as the initial condition and the inspiral
phase is computed for a finite duration. Consequently, the Fourier
spectrum for the low-frequency side (for $f \alt \Omega_0/\pi$)
becomes absent if we naively perform the Fourier transformation for
the numerical data.  To compensate the Fourier spectrum for the
low-frequency side, we combine a hypothetical waveform computed by the
Taylor-T4 formula, as often done (e.g., Refs.~\cite{BHBH12,ST08}). To
do so, we match the numerical waveforms with those by the Taylor-T4
formula at a time when the corresponding binary orbit in the numerical
simulation is approximately in a quasicircular state. As shown in
Fig.~\ref{FIG9}, two waveforms match well for a wide range of time, so
the resulting Fourier spectrum depends only weakly on the chosen
matching time.

Figure \ref{FIG12} plots $f h(f) D /m_0$ for various models.  In
Fig.~\ref{FIG12} (a)--(d), we also plot the Fourier spectra of a
gravitational waveform derived in the Taylor-T4 formula (long-dashed
curve) and by the Newtonian waveform (dashed curve) as (e.g., \cite{CF})
\beqn
f h(f) =\sqrt{{5 \over 24\pi}} {m_0 \over D} 
{Q^{1/2} \over 1+Q} (\pi m_0 f)^{-1/6}. \label{Nspec}
\eeqn
Here, ``Newtonian'' implies that the orbital motion is calculated in
the Newtonian plus 2.5 PN equations of motion, and the gravitational
waveform is computed by the quadrupole formula.

As Figure \ref{FIG12} indicates, the spectra of gravitational waves
emitted in the inspiral, merger, and ringdown phases are smoothly
connected. Nevertheless, we still see modulations of small amplitude
for $0.01 \alt f m_0 \alt 0.02$, which are likely to be caused by
slight modulations in the amplitude and/or phase of numerical
gravitational waveforms.

In the upper-horizontal and right-vertical axis, we plot the frequency
and averaged effective amplitude for hypothetical values $M_{\rm
NS}=1.35M_{\odot}$ and $D=100$ Mpc. Here, the averaged effective
amplitude is defined by the average of $f h(f)$ over the source
direction and the direction of the binary orbital plane as
\beqn
h_{\rm eff} & \equiv & 0.4 f h(f) \nonumber \\ 
&=& 9.6 \times 10^{-23} \biggl({f h(f) D/m_0 \over 0.1}\biggr)
\biggl({m_0 \over 5M_{\odot}}\biggr) \nonumber \\
&&~~~~~~~~~~~~~~~~ \times \biggl({D \over 100~{\rm Mpc}}\biggr)^{-1}. 
\label{heff}
\eeqn


Figure \ref{FIG12} shows that the spectrum shape has the following
universal qualitative feature: Irrespective of the values of ${\cal
C}$ and $Q$, $f h(f)$ (hereafter referred to as the spectrum
amplitude) decreases with the increase of $f$ and above a ``cut-off''
frequency, $f_{\rm cut}$, it decreases exponentially. For $f
\rightarrow 0$, $f h(f)$ is universally proportional to $f^{-1/6}$,
and $f < f_{\rm cut}$, $f h(f)$ is always written as $\propto
f^{-1/6}F(f)$, where $F(f)$ is a slowly varying function of $f$ and
the value is $\alt 1$. However, the detailed spectrum shape depends on
the values of ${\cal C}$ and $Q$, and Fig.~\ref{FIG12} exhibits a wide
variety of the possible spectrum shape as explained in the following.

For the case that the NS is tidally disrupted outside the ISCO, type I
gravitational waves are emitted (e.g., for model M20.145).  In this
case, the spectrum amplitude monotonically decreases, and above a
cut-off frequency, it exponentially decreases with
the increase of $f$. The cut-off frequency for model M20.145 is
$f_{\rm cut} \approx 0.04/m_0$. For $f \leq 0.03/m_0 \sim 0.8f_{\rm
cut}$, $F(f)$ may be approximately fitted by the 3PN formula for the
amplitude \cite{Kidder}, and thus, the spectrum may be described,
e.g., by the following way
\beqn 
h(f) =h_{\rm 3PN}(f) e^{-(f/f_{\rm cut})^{\sigma}}, \label{fit1} 
\eeqn 
where $h_{\rm 3PN}$ is the amplitude of gravitational waves derived in
the 3PN theory (see Eq. (79) of Ref.~\cite{Kidder}), and $\sigma$ is a
constant of order unity. The dotted curve in Fig.~\ref{FIG12} (a)
denotes the result of the fitting for $f_{\rm cut}=0.038/m_0$ and
$\sigma=2.2$, and we see that the fitting works well. For model 
M15.145, the fitting also works well and gives $f_{\rm cut}=0.030/m_0$ 
and $\sigma=2.2$. Because the NS is tidally disrupted farther outside 
the ISCO, the value of $f_{\rm cut}$ is smaller for model M15.145. 

The latest high-precision numerical study for the quasiequilibrium
states of the BH-NS binaries shows that the mass shedding of the NSs
occurs if the following condition is satisfied \cite{TBFS2}
\beqn
\Omega \geq \Omega_{\rm MS}
=C \biggl({GM_{\rm NS} \over R_{\rm NS}^3}\biggr)^{1/2}
\biggl(1+{M_{\rm NS} \over M_{\rm BH}}\biggr)^{1/2}, \label{MSEQ}
\eeqn
where the value of $C$ is $\approx 0.270$ for the $\Gamma=2$ polytropic
EOS. (Note that this is the relation which holds only for the NS of 
the irrotational velocity field and for the BH of no spin.) 
From this relation, we expect that the mass shedding sets in at a
frequency $f_{\rm MS}=\Omega_{\rm MS}/\pi$. We plot this expected 
frequency (0.0174, 0.0219, and $0.0265/m_0$) for models M20.145--M40.145
together with the Fourier spectra for models M20.145--M50.145 in
Fig.~\ref{FIG12} (e).


Figure \ref{FIG12} (a) and (e) show that any special feature is not
seen at $f=f_{\rm MS}$ in the spectrum even for the case that the NS
is tidally disrupted before the binary reaches the ISCO, e.g., for
model M20.145 (a small bump seen for $f < f_{\rm MS}$ is due to
modulation of the wave amplitude and wave frequency in the numerical
data).  Also, it is found that $f_{\rm cut}$ is much larger than
$f_{\rm MS}$, as already pointed out in Ref.~\cite{ST08}. This implies
that at $f=f_{\rm MS}$, the mass-shedding sets in for the NS, but the
NS is not tidally disrupted at such a low frequency. Even for the closer
orbits of $f > f_{\rm MS}$, the NS behaves as a self-gravitating star
although it transfers a small amount of mass to the companion BH, and
gravitational waves from this BH-NS binary are characterized basically
by the inspiral waveforms. However, because the tidal effect becomes
more important for the smaller orbital separation, the tidal
deformation of the NS is enhanced more with time, and eventually, the
tidal disruption occurs. At the tidal disruption, the gravitational
wave amplitude should quickly decrease, so it is natural to identify
the frequency of the last inspiral wave as $f_{\rm cut}$ in this case.

Here, we should emphasize that {\em the compactness (or radius) of the
NSs is reflected in $f_{\rm cut}$ not in $f_{\rm MS}$.} In the
existing idea, one naively assumes that the tidal disruption occurs at
$f=f_{\rm MS}$ and discusses that the radius of the NS (i.e., the EOS
of the NS) may be constrained by identifying the values of $f_{\rm 
MS}$ (e.g. Ref.~\cite{valli}). However, our present result indicates 
that this possibility is unlikely.  To constrain the EOS of the NS from 
gravitational waves detected, we have to theoretically determine  
$f_{\rm cut}$ for a variety of the EOSs and masses of the BH and NS. 

For models M20.160 and M30.145 [see Fig.~\ref{FIG12} (a) and (b)], the
spectrum shape is similar to that for model M20.145, but the cut-off
frequency is higher as $f_{\rm cut} \sim 0.06/m_0$ (see below). This
cut-off frequency is still lower than the frequency of the fundamental
quasinormal mode ($f_{\rm QNM}$; see the arrow of Fig.~\ref{FIG12} (a)
and (b)). This indicates that the tidal disruption occurs at an
orbit corresponding to $f=f_{\rm cut}$.  This cut-off frequency, which
is higher than that for model M20.145, reflects the fact that the
tidal disruption occurs at a closer orbit than that for model M20.145
due to the larger compactness or mass ratio of the system. The cut-off
frequency appears to be much higher than the frequency of
gravitational waves emitted at the ISCO, which is $f_{\rm lso} \sim
0.03/m_0$. This implies that gravitational waves of $f_{\rm lso} \alt
f \alt f_{\rm cut}$ are not emitted by the inspiral motion of the
binary but by a motion in the merger phase. (This is the reason why we
do not classify the waveforms for models M20.160 and M30.145 into type I
but into type II.)  In this case, it is not appropriate to fit the
spectrum by Eq. (\ref{fit1}) but by a different function (see below). 

Another interesting feature found in the spectrum of model M30.145 is
that no peak associated with the quasinormal mode appears at $f=f_{\rm
QNM}$ (see arrow in Fig.~\ref{FIG12} (b)). The reason is simply that
the amplitude of the ringdown waveform is much smaller than that of
the merger waveform, as pointed out in Sec.~\ref{sec:gw}.

For models M40.145, M50.145, and M30.178 (see Fig.~\ref{FIG12} (c),
(d), and (f)), the gravitational waveforms are type III. In these
cases, the spectrum amplitude also steeply decreases above a cut-off
frequency, but the feature of the spectrum shape is qualitatively
different from those of models M20.145, M30.145, and M20.160, because
the gradual decrease of the spectrum amplitude continues approximately up
to the frequency of the quasinormal mode (see the arrows in these
figures). Namely, $f_{\rm cut}$ is approximately equal to $f_{\rm
QNM}$, and thus, the cut-off frequency does not indicate that 
the tidal disruption occurs at an orbit of $f=f_{\rm cut}$. In other
words, the signal of the tidal disruption is absent in these cases.

Another difference is seen in the spectrum shape for a frequency
slightly smaller than $f_{\rm cut}$. For models M20.145, M30.145, and
M20.160, a spectral index, $n \equiv -d\ln(fh(f))/d\ln(f)$, for $f
\alt f_{\rm cut}$ monotonically increases with $f$. For models
M40.145, M50.145, and M30.178, by contrast, the value of $n$ slightly
decreases with $f$, and a ``plateau'' appears.  This spectrum shape is
similar to that for the merger of unequal-mass BH-BH binaries
\cite{BBBB}.  This is reasonable because in this case, tidal effects
do not play an important role during the inspiral and merger phases,
and the merger process should be qualitatively the same as that in
BH-BH binaries.

In Fig.~\ref{FIG12} (f), we compare the spectrum shape for models
M30.145, M30.160, and M30.178, for which the masses of the BH and NS
are identical whereas the compactness of the NS is different. This
figure illustrates that for the larger compactness of the NS, the cut-off
frequency is higher. Also, the width of the plateau is larger for the
more compact models (M30.160 and M30.178).  As discussed above, these
differences reflect the difference in the evolution process of the
late inspiral and merger phases. Because the spectrum shape near
$f=f_{\rm cut}$ is significantly different among three models,
observing such high-frequency gravitational waves will play a special
role in constraining the compactness (i.e., radius) of the NS.


To systematically identify the cut-off frequency, the fitting of the
spectrum with a specific function is useful, as illustrated for model
M20.145. The spectra associated with the inspiral phase for models
M30.145--M50.145, M20.160, M20.178, M30.160, and M30.178 are also fitted 
by the 3PN amplitude, $h_{\rm 3PN}(f)$, for $f \alt f_{\rm lso} \sim
0.03/m_0$.  However, this should not be the case for $f_{\rm lso} \alt
f \alt f_{\rm cut}$ because gravitational waves for this
high-frequency component are not emitted by the inspiral motion, but
by a motion of the material associated with the merger and infalling
into the BH. Hence, the spectrum should not be fitted by 
Eq. (\ref{fit1}) but by
\beqn
h(f) =h_{\rm 3PN}(f) e^{-(f/f_{\rm ins})^{\sigma}}
+h_{\rm merger}(f), 
\label{fit2}
\eeqn
where $h_{\rm merger}(f)$ denotes the spectrum of gravitational 
waves emitted in the merger phase, and $f_{\rm ins}$ is a 
frequency of 0.01--$0.03/m_0$. In this paper, we fix $\sigma=3.5$ 
to reduce the total number of free parameters, and 
choose the following function for $h_{\rm merger}(f)$, 
\beqn
h_{\rm merger}(f)={A m_0 \over D f} e^{-(f/f_{\rm cut})^{\sigma_{\rm cut}}}
[1-e^{-(f/f_{\rm ins2})^5}], 
\eeqn
where $A$ and $\sigma_{\rm cut}$ are free parameters. We add 
a free parameter $f_{\rm ins2}$ for which the value is close 
to $f_{\rm ins}$, because the fitting is achieved in a better manner 
with it. 

The dotted curves in Fig.~\ref{FIG12}~(b)--(d) denote the results for
$(f_{\rm ins}m_0,f_{\rm ins2}m_0, A, f_{\rm cut}m_0, \sigma_{\rm
cut})=$(0.014, 0.014, 0.130, 0.063, 2.90) for model M30.145, (0.016,
0.016, 0.103, 0.079, 4.60) for model M40.145, and (0.019, 0.020,
0.090, 0.087, 3.70) for model M50.145.  As Fig.~\ref{FIG12} shows, the
spectrum is well fitted by this simple fitting function.

The fitting for model M50.145 (and M30.178) is not as excellent as
those for other models. In these cases, the inclination of the plateau
is rather flat in the high-frequency region.  This feature is
universal for the gravitational waveforms emitted in the merger of
BH-BH binaries.  Thus, in such a case, another fitting function
proposed, e.g. in Ref.~\cite{BBBB}, may be better.  The fitting
procedure here is, however, still robust in extracting important
physical information, as discussed in the following.

\subsection{Cut-off frequency and determining the compactness of the NS}

\begin{figure}[t]
\begin{center}
\epsfxsize=3.3in
\leavevmode
\epsffile{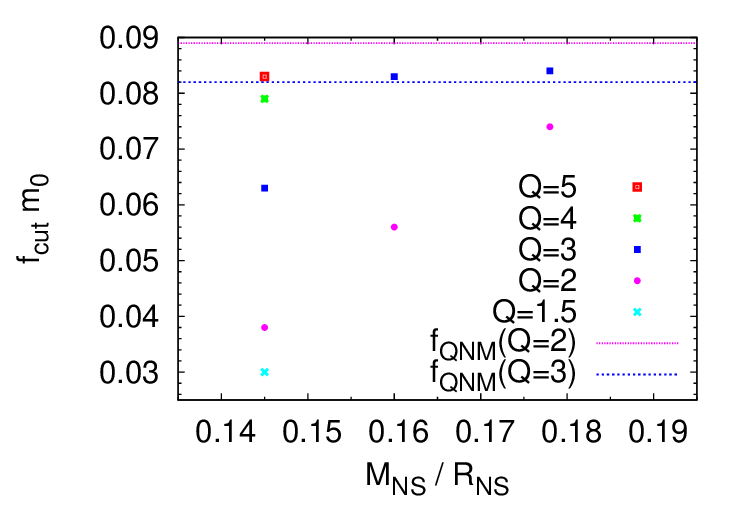}
\end{center}
\vspace{-7mm}
\caption{$f_{\rm cut}$ as a function of ${\cal C}$ for various values
of $Q$. The dashed and dotted lines denote the value of $f_{\rm QNM}$
for $Q=2$ and 3, respectively.  For $Q=4$ and 5, $f_{\rm QNM}$ is
slightly smaller than that for $Q=3$, and $\approx 0.078/m_0$ and
$0.075/m_0$, respectively.
\label{FIG13}}
\end{figure}

In the fitting procedure described in the previous subsections, the
most important output is the cut-off frequency, $f_{\rm cut}$, in
which information about the compactness (or radius) of the NS may be
reflected. For the case that the NS is swallowed by the BH with no
tidal disruption (e.g., $Q \agt 4$), it approximately indicates the
frequency of the quasinormal mode of the formed BH, and thus, it will
not be possible to extract the compactness of the NSs from the
gravitational-wave signal.  By contrast, for the case that the NS is
tidally disrupted before it is swallowed by the BH (i.e., for the
small values of $Q$ and ${\cal C}$), $f_{\rm cut}$ is much smaller
than $f_{\rm QNM}$ and indicates a characteristic frequency of
gravitational waves emitted at the tidal disruption (see
Fig. \ref{FIG13}). We find that for models M20.145, M20.160, M20.178,
M30.145, and M30.160, $f_{\rm cut} m_0 \approx 0.038$, 0.056, 0.070, 
0.063, and 0.083.  For $Q=2$, the value of $f_{\rm cut}$ is smaller 
than the frequency of the quasinormal mode, $f_{\rm QNM} \sim 
0.09/m_0$ for a wide value of ${\cal C}$.  This indicates that we will 
be able to constrain the compactness of the NS if we detect
gravitational waves in the merger phase for a binary composed of a
low-mass BH and an NS with $Q \alt 2$.  For model M30.160, $f_{\rm
cut}$ is approximately equal to $f_{\rm QNM}$, whereas $f_{\rm cut} <
f_{\rm QNM}$ for mode M30.145.  Thus, if the compactness of the NS is
fairly small ${\cal C} < 0.16$, we will have a chance that the
detection of gravitational waves from a binary of $Q \approx 3$
constrains the compactness of the NS. By contrast for $Q \geq 4$,
gravitational waves are not likely to have robust information about
the compactness of the NSs.

Figure \ref{FIG12} (a)--(d) and (f) show that the cut-off frequency is
$f=f_{\rm cut} \sim 1.4$--2.2 kHz and the effective amplitude at
$f=f_{\rm cut}$ is universally $\sim 1 \times 10^{-22}$ for
hypothetical values $D=100$ Mpc and $M_{\rm NS}=1.35M_{\odot}$. Even
for the optimistic direction of the source with respect to the plane
of the detector's arm and of its binary orbital plane, the effective
amplitude is at most $\approx 2.5 \times 10^{-22}$. The designed
sensitivity of the advanced LIGO is $\leq 3 \times 10^{-22}$ at $f
\geq 1$ kHz \cite{KIP}. This implies that it will not be able to
detect gravitational waves during the merger and ringdown phases by
the laserinterferometric detectors of standard design for $D \agt 100$
Mpc. To detect gravitational waves at such high frequency, the
detectors composed of special instrument (e.g., resonant-side band
extraction), which has a high sensitivity for high-frequency
gravitational waves, is necessary. As mentioned above, gravitational
waves for $f \alt f_{\rm cut}$ carry important information of the NS
radius. It is strongly desired that a detector which is sensitive for
the frequency of 1 kHz $\alt f \alt 3$ kHz will be developed in the
future.

\subsection{Kick velocity}

\begin{figure}[thb]
\vspace{-4mm}
\begin{center}
\epsfxsize=3.2in
\leavevmode
\epsffile{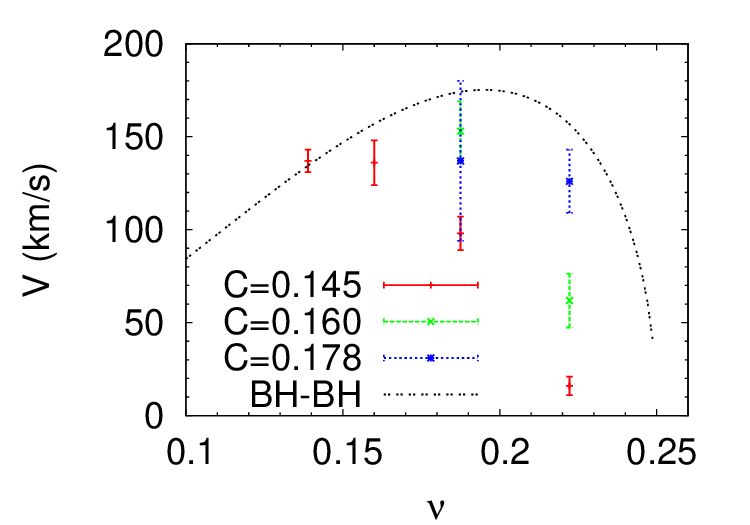}
\end{center}
\vspace{-5mm}
\caption{Kick velocity as a function of $\nu \equiv Q/(1+Q)^2$ for
different compactness of the NSs. The dot-dotted curve denotes the
fitting formula derived in Ref.~\cite{kick2} for the merger of
nonspinning BH-BH binaries.
\label{FIG14}}
\end{figure}

Kick velocity induced by anisotropic gravitational radiation is
estimated from the total linear momentum carried by gravitational
waves. Figure \ref{FIG14} plots the kick velocity as a function of the 
ratio of reduced mass to total mass for different compactness of 
the NSs (see also Table IV). 

We find that the kick velocity depends strongly on the mass ratio and
compactness of the NS. More specifically, it depends strongly on whether
the NS is tidally disrupted or not. If the tidal disruption does not
occur, the NS simply falls into the companion BH in a basic picture
(although tidal deformation and strong elongation of the NS still
occur just before it falls into the BH). In such a case, the merger
process is similar to that in the merger of nonspinning BH-BH
binaries, and the kick velocity should agree approximately with the
results for those systems. To confirm this fact, we compare
our numerical results with the fitting formula derived in
Ref.~\cite{kick2} (see dot-dotted curve in Fig.~\ref{FIG14}).  Figure
\ref{FIG14} shows that our results for models M40.145, M50.145,
M30.160, and M30.178 are in a reasonable agreement with the fitting
formula of Ref.~\cite{kick2}; the kick velocity is $\sim 100$--200
km/s.  This indicates that our numerical results are reliable.

In the case that the NS is tidally disrupted, the kick velocity is
significantly suppressed.  The primary reason is that the kick is
excited most efficiently when two stars merge; more specifically, when
the amplitude of gravitational waves becomes maximum. The maximum
amplitude of inspiral gravitational waves is higher for the case
that two stars have more compact orbits (i.e., the NS is more
compact). Also, the amplitude of gravitational waves in the merger
phase becomes higher when an NS falls into a companion BH without
tidal deformation. For the case that the tidal disruption happens,
close inspiral orbits are prohibited, and as a result, the enhancement
of the gravitational-wave amplitude is suppressed as we showed in Sec.
\ref{sec:gw}. Due to this fact, the kick velocity is smaller for the
smaller values of compactness and mass ratio. 

\section{Summary}

New general relativistic simulation for the inspiral, merger, and
ringdown of BH-NS binaries are performed by a new AMR code {\tt
SACRA}.  This paper presents the numerical results of a longterm
simulation for a variety of mass ratio and compactness of the NS for
the first time. For the simulation, we adopt initial conditions for
which the unphysical eccentricity is much smaller than that in our
previous study \cite{SU06,ST08}, and also the initial orbital
separation is much larger. In the present results, the binaries spend
in the inspiral phase for 4--7 orbits, and in the last several orbits,
the eccentricity is sufficiently suppressed to be $\sim 0.01$ and the
orbit becomes approximately quasicircular. This is reflected in
successful computation of gravitational waveforms for the inspiral
phase which agree well with those predicted by the PN theory.  Because
the moving-puncture approach is adopted, the merger and subsequent
evolution of the BH with the surrounding material are simulated for a
long time until the system relaxes to a quasisteady state. As a
result, accurate gravitational waveforms for the late inspiral,
merger, and ringdown phases are derived successfully.


By the present self-consistent simulation, we reconfirm the finding in
our previous paper \cite{ST08} that the merger process depends
sensitively on the mass ratio, $Q$, and compactness of the NS, ${\cal
C}$. Only for the case that both $Q$ and ${\cal C}$ are sufficiently
small, the NS is tidally disrupted by the companion BH. For example,
for the chosen EOS in this paper (polytropic EOS with $\Gamma=2$), the
NS of ${\cal C}=0.145$ is tidally disrupted only for $Q \alt 3$ before
the binary reaches the ISCO.  By analyzing the spectrum of
gravitational waves, we confirm that even in the case that the NS
satisfies the condition for the onset of the mass shedding [see
Eq. (\ref{MSEQ})], the tidal disruption does not occur immediately.
Rather, the NS behaves as a self-gravitating star for a while during
the mass-shedding phase. Therefore, the tidal disruption occurs for
the more restricted case than the mass shedding does.  This fact also
implies that for determining the condition of tidal disruption,
numerical-relativity simulation is the unique approach.


The parameter space for the formation of a long-lived disk surrounding
the BH is even more restricted. For ${\cal C}=0.145$, the disk of
lifetime $\gg 10$ ms is formed only for $Q \alt 2$, and for ${\cal
C}=0.160$, the disk mass is at most $10^{-3}M_*$ even for $Q=2$. The
mass of the formed disk is $\agt 0.01$--$0.02M_*$ for $Q=2$--1.5 even for
relatively less compact NSs with ${\cal C}=0.145$, and thus, the disk
is not very massive, even if it is formed. This conclusion is
consistent with the results of Ref.~\cite{CORNELL}.

At present, the precise EOS of high-density nuclear matter is 
unknown \cite{LP}.  However, many EOSs predict that the radius of NSs
of canonical mass 1.25--$1.45M_{\odot}$ is smaller than $\sim 12$ km
(e.g., Ref.~\cite{EOS}). Namely, the compactness of the NS is larger than
0.155. If the EOS is really stiff and the NS is compact as the nuclear
theory predicts, the disk will not be formed after the merger between 
nonspinning BH and NS, as pointed out by Miller \cite{CMiller}. 
(However, this will not be the case if the BH has the spin of 
substantial magnitude.) 


We find that gravitational waveforms from BH-NS binaries are roughly
classified into the following three types. (i) When the NS is tidally
disrupted during the inspiral phase, the gravitational waveform is
characterized by the inspiral waveform and subsequent sudden
shut-down. In this case, the amplitude of the Fourier spectrum
monotonically decreases with the increase of the frequency, and at a
cut-off frequency $f_{\rm cut}$, which is the frequency of
gravitational waves when the NS is tidally disrupted (not equal to
$f_{\rm MS}$), the spectrum amplitude decreases exponentially [see
Eqs.~(\ref{fit1}) and (\ref{fit2})]. We refer to this waveform as type
I in this paper.  (ii) In the case that the NS is tidally disrupted
but most of the material falls into the companion BH before the tidal
disruption is completed, the gravitational waveform is characterized
by the inspiral and merger waveforms. In this case, ringdown
gravitational waves are excited in the final phase, but its amplitude
is much smaller than those of late inspiral and merger
gravitational waves. As a result, the shape of the Fourier spectrum
is similar to that of type I but the cut-off frequency does not
correspond to the frequency of the last inspiral orbit nor the
frequency of the quasinormal mode.  We refer to this waveform as type
II.  (iii) When the NS is not tidally disrupted before it is swallowed
by the BH, the gravitational waveform is characterized by the
inspiral, merger, and ringdown waveforms.  In this case, the amplitude
of the Fourier spectrum monotonically decreases with the increase of
the frequency in the inspiral phase as in the cases (i) and
(ii). However, in the late inspiral and merger phases, the
gravitational-wave amplitude increases and, as a result, a plateau
appears in the Fourier spectrum of $f \alt f_{\rm cut}$.  Then the
spectrum amplitude exponentially decreases for $f > f_{\rm cut}$ [see
Eq. (\ref{fit2})].  In this case, $f_{\rm cut}$ is approximately equal
to the frequency of the fundamental quasinormal mode of the formed BH,
$f_{\rm QNM}$, and does not have any information about tidal
disruption (and thus compactness of the NS). We refer to this waveform
as type III.

We fit the spectrum of gravitational waves with hypothetical functions
of a small number of parameters. We find that the fitting works well
irrespective of the values of $Q$ and ${\cal C}$. By this fitting, we
systematically determine the value of the cut-off frequency, $f_{\rm
cut}$ and confirm the following fact: For the case that the NS is not
tidally disrupted, the value of $f_{\rm cut}$ is approximately equal
to the value of $f_{\rm QNM}$, as mentioned above. By contrast, for
the case that the tidal disruption occurs, the value of $f_{\rm cut}$
is smaller than $f_{\rm QNM}$, and it depends strongly on $Q$ and
${\cal C}$. The tidal disruption occurs for a wide range of ${\cal C}$
if $Q$ is smaller than $\sim 2$ . Thus, if gravitational waves from
binaries composed of a low-mass BH and NS at tidal disruption are
observed, it may be possible to constrain the compactness of the NS,
and as a result, the EOS of the NS.

We also reconfirm that the frequency at the onset of the mass shedding of
the NSs is not reflected in the spectrum in an outstanding manner.
Namely, the compactness (or radius) of the NSs is reflected in $f_{\rm
cut}$ not in $f_{\rm MS}$. In the existing idea, one naively assumes
that tidal disruption occurs at $f=f_{\rm MS}$ and discusses that the
radius of the NS (i.e., the EOS of the NS) may be constrained by
identifying the values of $f_{\rm MS}$
(e.g. Ref.~\cite{valli}). However, our present result indicates that
this possibility is unlikely.  To constrain the EOS of the NS from
gravitational waves detected, we have to determine $f_{\rm cut}$ for a
variety of the EOSs and masses of the BH and NS, that can be done only
by numerical-relativity simulation.

Kick velocity induced by asymmetric gravitational wave emission is
also computed. When tidal disruption does not occur (e.g., for $Q=5$),
our result agrees approximately with that derived for the merger of
nonspinning BH-BH binaries; the kick velocity is $\sim 100$--200 km/s.
For the case that the tidal disruption occurs, the kick velocity is
significantly suppressed, and it is much smaller than 100 km/s.  This
is due to the fact that gravitational wave amplitude in the last
inspiral and merger phases is suppressed.

As shown in this paper, gravitational waveforms in the merger and
ringdown phases depend strongly on the compactness of the NS and mass
ratio of the binary.  This results primarily from the fact that the
degree of the tidal deformation and condition for the onset of the mass
shedding and tidal disruption depend strongly on these two
parameters. Gravitational waveforms should also depend on the EOS of
the NS and on the spin of the BH, as illustrated, e.g. in
Ref.~\cite{ISM}. Thus, in the subsequent work, we will systematically
perform the simulation for a variety of the EOSs of the NS and BH spin,
to clarify the dependence of the gravitational waveform and its
spectrum on these additional parameters. 


\begin{acknowledgments}
All the initial conditions for the present numerical simulation are
computed using the free library LORENE \cite{LORENE}.  We thank
members in the Meudon Relativity Group, in particular, Eric
Gourgoulhon, for developing LORENE.  We also thank Alessandra Buonanno
for telling methods for the analysis of numerical gravitational
waveforms.  The numerical computations were in part performed on
NEC-SX8 at Yukawa Institute of Theoretical Physics of Kyoto
University.  This work was supported by Monbukagakusho Grant
No. 19540263 and 20105004.
\end{acknowledgments}

\section{Erratum and addendum}

(The original version of this article was submitted on Feb 3, 2009. This
erratum is added on Jul 23, 2012.)

In the original version of this article, we reported numerical results
obtained by our {\tt SACRA} code employing a simple $\Gamma$-law
equation of state with $\Gamma=2$.  During subsequent
studies~\cite{kst2010,kost2011}, we noticed that we systematically {\it
underestimated} disk masses in the original work. The reason is that we
evolved hydrodynamic variables and estimated disk masses only in a
domain of the size $\sim 200^3 \; \mathrm{km}^3$, although Einstein's
equations were solved in domains of size $\sim 1500^3$--$2000^3\;
\mathrm{km}^3$. A small-size domain for hydrodynamics is insufficient
for the estimation of the disk mass, if tidal disruption occurs.  In
particular, for a large NS radius, the tidally disrupted material
extends far away from the central region, and the disk mass was
significantly underestimated. This problem was found during the writing
of~\cite{kost2011}, and the erratum of~\cite{kst2010,kst2010e} were
already published.

For this reason, we reperformed simulations for the same models as those
of the original version, enlarging the computational domain of
hydrodynamics. To estimate disk masses more accurately, in addition, we
improved the grid resolutions from $N \le 36$ to $N=40$, 50, and 60;
with $N=60$, the diameters of NSs are covered by $\approx 100$ grid
points, and the diameters of apparent horizons of BHs are by $\approx
20Q$ grid points. With this improved grid resolution, we also reanalyzed
gravitational waves. In addition, the fitting formula for the
quasinormal-mode frequency is updated to the latest, more sophisticated
one derived in~\cite{bcs2009}.

\begin{figure}[tbp]
 \includegraphics[width=85mm,clip]{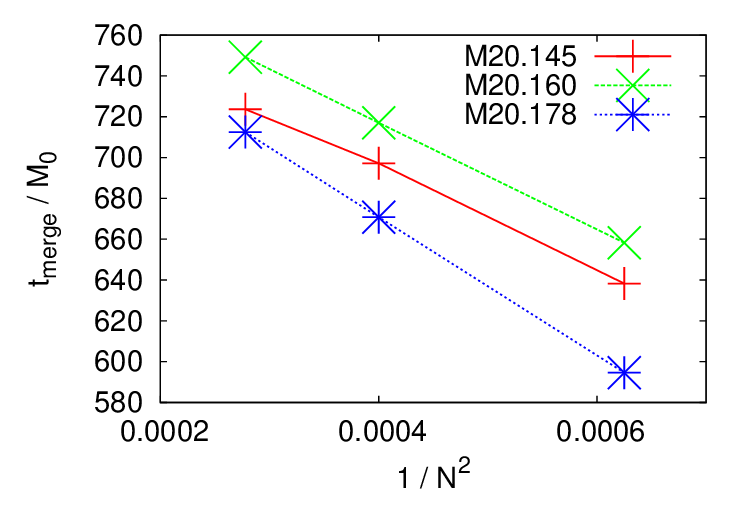}
 \vspace{-4mm}
 \caption{The merger time normalized by the Arnowitt-Deser-Misner mass
 of the initial configuration for models with $Q=2$ as a function of
 $1/N^2$. The left, middle, and right points correspond to the results
 of $N=60$, 50, and 40, respectively.} \label{fig:mergertime}
\end{figure}

First, we show the convergence properties of our updated simulations.
Figure~\ref{fig:mergertime} plots the merger time for $Q=2$ models as
a function of $1/N^2(\propto \Delta x^2)$, and indicates that it
converges approximately at the second order, which is expected for
{\tt SACRA}. Thus, a reasonable convergence is indicated to be
achieved in our new simulations. The convergence properties of other
quantities are discussed below.

\begin{table*}[tbp]
 \caption{Corrected numerical results for the remnants: Compare with
 Table III.}
 \begin{tabular}{cccccccccc} \hline \hline Model &
 $M_{r>r_\mathrm{AH}}/M_*$ & $M_\mathrm{BH,f}/M_0$ & $C_e / 4\pi
 M_0$ & $M_\mathrm{irr} / M_0$ & $C_p / C_e$ & $a_\mathrm{f1}$ &
 $a_\mathrm{f2}$ & $a_\mathrm{f3}$ & $f_\mathrm{cut} m_0$ \\
  \hline \hline
  M20.145 ($N=60$) & 0.060 & 0.970 & 0.970 & 0.903 & 0.898 & ~0.692~ &
                              0.682~ & ~0.683~ & ~0.036~ \\
  M20.145 ($N=50$) & 0.065 & 0.969 & 0.968 & 0.902 & 0.899 & 0.693 &
                              0.679 & 0.680 & 0.037 \\
  M20.145 ($N=40$) & 0.066 & 0.969 & 0.969 & 0.901 & 0.898 & 0.703 &
                              0.682 & 0.680 & 0.034 \\
  \hline
  M20.160 ($N=60$) & 0.021 & 0.981 & 0.980 & 0.912 & 0.897 & 0.694 &
                              0.684 & 0.685 & 0.055 \\
  M20.160 ($N=50$) & 0.025 & 0.980 & 0.980 & 0.911 & 0.897 & 0.697 &
                              0.684 & 0.685 & 0.054 \\
  M20.160 ($N=40$) & 0.028 & 0.980 & 0.979 & 0.910 & 0.897 & 0.704 &
                              0.684 & 0.685 & 0.052 \\
  \hline
  M20.178 ($N=60$) & 0.0021 & 0.983 & 0.982 & 0.917 & 0.903 & 0.680 &
                              0.667 & 0.667 & 0.078 \\
  M20.178 ($N=50$) & 0.0027 & 0.984 & 0.982 & 0.917 & 0.905 & 0.688 &
                              0.666 & 0.662 & 0.075 \\
  M20.178 ($N=40$) & 0.0067 & 0.984 & 0.982 & 0.916 & 0.902 & 0.701 &
                              0.670 & 0.662 & 0.069 \\
  \hline
  M30.145 ($N=60$) & 0.044 & 0.978 & 0.978 & 0.936 & 0.936 & 0.566 &
                              0.556 & 0.557 & 0.065 \\
  M30.145 ($N=50$) & 0.047 & 0.978 & 0.978 & 0.935 & 0.936 & 0.585 &
                              0.557 & 0.558 & 0.063 \\
  M30.145 ($N=40$) & 0.055 & 0.977 & 0.976 & 0.934 & 0.936 & 0.573 &
                              0.556 & 0.558 & 0.056 \\
  \hline
  M30.160 ($N=60$) & 0.014 & 0.982 & 0.982 & 0.940 & 0.937 & 0.565 &
                              0.554 & 0.555 & 0.080 \\
  M30.160 ($N=50$) & 0.017 & 0.982 & 0.981 & 0.939 & 0.937 & 0.569 &
                              0.554 & 0.555 & 0.079 \\
  M30.160 ($N=40$) & 0.023 & 0.982 & 0.981 & 0.939 & 0.936 & 0.578 &
                              0.556 & 0.555 & 0.075 \\
  \hline
  M30.178 ($N=60$) & 0.0011 & 0.983 & 0.982 & 0.941 & 0.938 & 0.561 &
                              0.547 & 0.548 & 0.086 \\
  M30.178 ($N=50$) & 0.0020 & 0.983 & 0.982 & 0.941 & 0.938 & 0.568 &
                              0.549 & 0.550 & 0.086 \\
  M30.178 ($N=40$) & 0.0039 & 0.984 & 0.983 & 0.941 & 0.937 & 0.580 &
                              0.551 & 0.550 & 0.083 \\
  \hline
  M40.145 ($N=60$) & 0.016 & 0.986 & 0.985 & 0.955 & 0.955 & 0.485 &
                              0.475 & 0.477 & 0.080 \\
  M40.145 ($N=50$) & 0.018 & 0.985 & 0.985 & 0.955 & 0.955 & 0.486 &
                              0.475 & 0.476 & 0.079 \\
  M40.145 ($N=40$) & 0.026 & 0.985 & 0.984 & 0.954 & 0.955 & 0.492 &
                              0.474 & 0.476 & 0.076 \\
  \hline
  M50.145 ($N=60$) & $4 \times 10^{-4}$ & 0.990 & 0.989 & 0.966 & 0.966
                      & 0.426 & 0.417 & 0.418 & 0.082 \\
  M50.145 ($N=50$) & $7 \times 10^{-4}$ & 0.990 & 0.989 & 0.967 & 0.965
                      & 0.431 & 0.418 & 0.420 & 0.082 \\
  M50.145 ($N=40$) & 0.0022 & 0.991 & 0.990 & 0.966 & 0.965 & 0.435 &
                              0.419 & 0.420 & 0.080 \\
 \hline \hline
 \end{tabular}
 \label{table:remnant}
\end{table*}

\begin{figure}[tb]
 \includegraphics[width=85mm,clip]{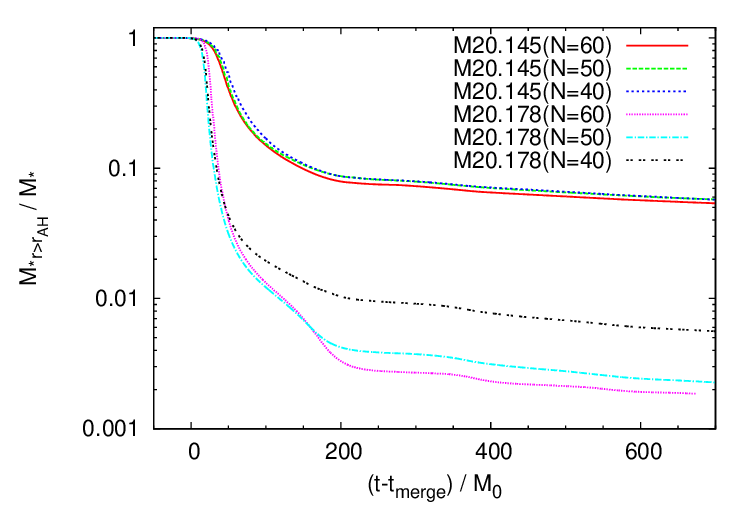}
 \vspace{-5mm}
 \caption{Evolution of the rest mass of the material located outside the
 apparent horizon for different grid resolutions of models M20.145 and
 M20.178.} \label{fig:diskconv}
\end{figure}

The corrected quantities for the merger remnant are listed in
Table~\ref{table:remnant} for three grid resolutions. This shows that an
approximate convergence of the disk mass is achieved for cases
$M_{r>r_\mathrm{AH}} \gtrsim 0.01 M_*$, and the convergence appears to
be not so good when $M_{r>r_\mathrm{AH}} \lesssim 0.01 M_*$. This is
simply because such a small disk mass is supplied from a lower-density
region near the surface of NSs for which an accurate computation is
forbidden in a poor-resolution run; i.e., the steep density gradients
near the surface cannot be accurately resolved, leading to a slight
spurious expansion of the surface which spuriously increases the
resulting disk mass. We estimate that the disk mass is overestimated
even in the run with $N=60$ by a factor of 2 -- 3, when the resulting
disk mass is of order $10^{-3} M_*$ or less. Figure~\ref{fig:diskconv}
plots the evolution of $M_{r>r_\mathrm{AH}}$ for three grid
resolutions. This indeed indicates that the convergence is good when
$M_{r>r_\mathrm{AH}} \gtrsim 0.01 M_*$ but not as good when
$M_{r>r_\mathrm{AH}} \sim 10^{-3} M_*$. Although the remnant disk mass
may be overestimated by $10^{-3} M_*$ -- $10^{-2}M_*$, this error is
acceptable for the problem of disk formation with $\sim 0.1 M_*$. We
also note that our results of the disk mass for models M30.145 are in
approximate agreement with those of numerical-relativity simulations
performed by other two groups~\cite{dfkot2010,elsb2009}.

Table \ref{table:remnant} shows that corrected quantities of the remnant
BH are in the convergent regime within $O(0.1) \%$ errors except for
$a_\mathrm{f1}$, indicating a reasonable convergence. The reason for the
poorer convergence of $a_\mathrm{f1}$ is due to its dependence on the
mass and angular momentum of the remnant disk. However, we find that the
disagreement between $a_\mathrm{f1}$ and other two, $a_\mathrm{f2}$ and
$a_\mathrm{f3}$, becomes smaller as the grid resolution is improved, and
the error is $\lesssim 0.015$ for $N=60$. We note that $M_\mathrm{BH,f}$
and $C_e / 4\pi$ always agree approximately with each other irrespective
of $N$.

\begin{figure*}[tbp]
 \begin{tabular}{ccc}
  \includegraphics[width=50mm,clip]{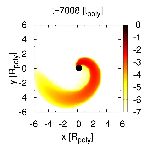} &
  \includegraphics[width=50mm,clip]{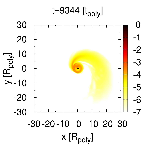} &
  \includegraphics[width=50mm,clip]{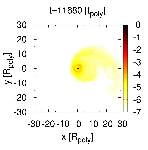}
 \end{tabular}
 \vspace{-5mm}
 \caption{Evolution of the rest-mass density profile in the logarithmic
 scale and the location of apparent horizons (black filled circles) for
 model M30.145 (N=60). All the quantities are shown in the polytropic
 unit, in which the unit length is $\approx 14$ km for a hypothetical NS
 mass $M_\mathrm{NS} = 1.35 M_\odot$.} \label{fig:snap}
\end{figure*}

\begin{figure*}[tbp]
 \begin{tabular}{cc}
  \includegraphics[width=80mm,clip]{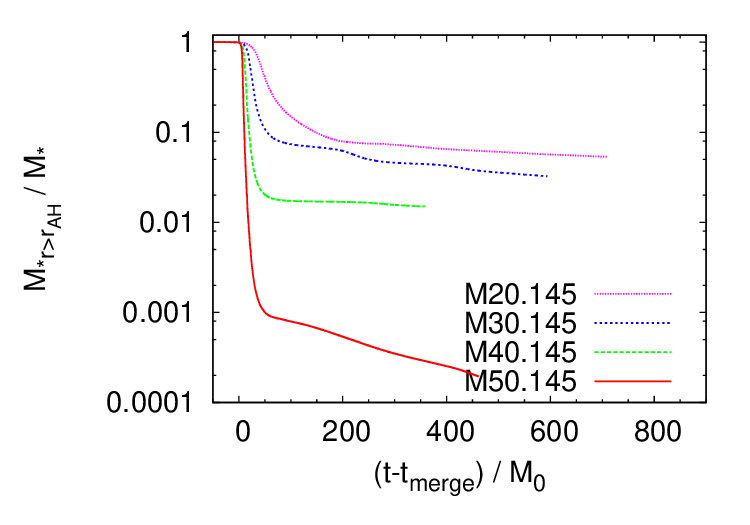} &
  \includegraphics[width=80mm,clip]{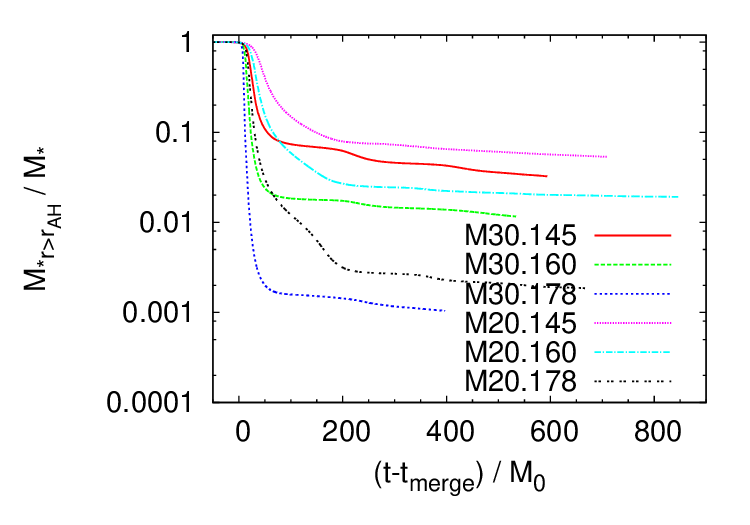}
 \end{tabular}
 \vspace{-5mm}
 \caption{Left: Evolution of the rest mass of the material located
 outside apparent horizons for models M20.145, M30.145, M40.145, and
 M50.145 in $N=60$ runs. Right: The same as the left panel, but for
 models M30.145, M30.160, M30.178, M20.145, M20.160, M20.178 in $N=60$
 runs. For a hypothetical NS mass $M_\mathrm{NS} = 1.35 M_\odot$, $100
 M_0 \approx 2.00 (Q+1)/3$~ms. } \label{fig:disk}
\end{figure*}

The corrected remnant disk mass for $N=60$ is significantly larger than
that in our previous simulations. The reason for the mistake in the
original version is due to the fact that we did not follow the motion of
the material which is ejected by tidal disruption and escapes outside
the central domain of the size $\sim 200^3 \; \mathrm{km}^3$, even when
they are bound and supposed to eventually return to the neighborhood of
the BH. The new {\tt SACRA} can correctly estimate the mass of the
returned material. Snapshots of the rest-mass density profiles for model
M30.145 are displayed in Fig.~\ref{fig:snap}, which shows that the
hydrodynamics processes are solved appropriately in the far
region. Figure~\ref{fig:disk} plots the time evolution of
$M_{r>r_\mathrm{AH}}$ for all the models with $N=60$. It is found that
the remnant disk mass after $\approx 10$~ms of the merger exceeds $0.01
M_\odot$ for a wider range of binary parameters (for a hypothetical NS
mass $1.35M_{\odot}$). However, the remnant disk mass is still small for
a large mass ratio, $Q = 5$ or for a compact NS, $\mathcal{C} = 0.178$
for which the tidal effect plays a minor role in the merger process.

By contrast, quantities associated with the remnant BH do not differ
much from those obtained in the original version. This fact suggests
that the disrupted material does not affect the remnant BH
strongly. This is reasonable, because the mass of the remnant BH is
always much larger than that of the remnant disk.

\begin{figure*}[tbp]
 \begin{tabular}{cc}
  \includegraphics[width=80mm,clip]{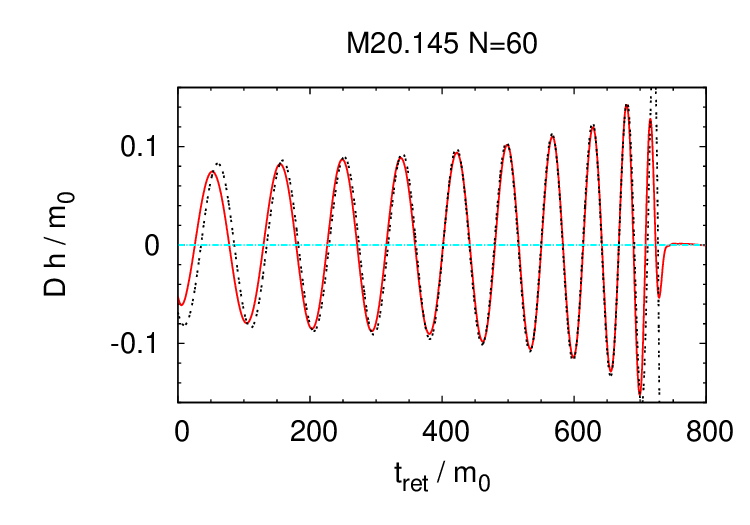} &
  \includegraphics[width=80mm,clip]{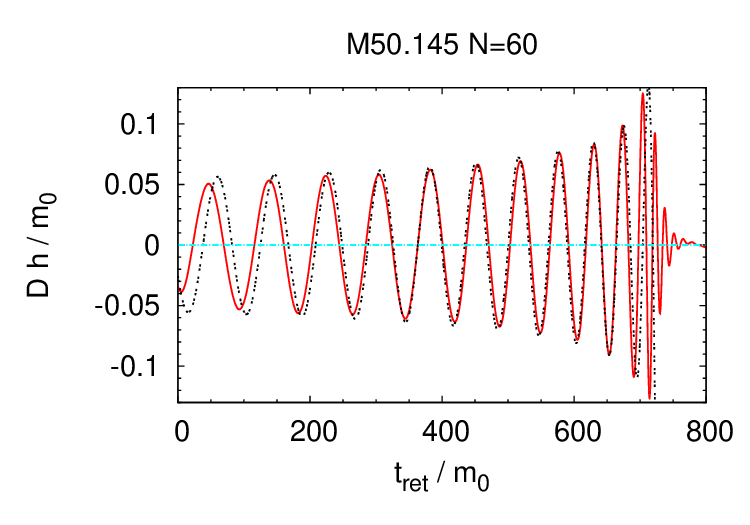}
 \end{tabular}
 \vspace{-5mm}
 \caption{Gravitational waveforms observed along the {\it z} axis (solid
 curve) for models M20.145 (left) and M50.145 (right). $t_\mathrm{ret}$
 denotes the retarded time and $m_0$ is the total mass. The solid
 curves denote the results obtained in $N=60$ runs and the dot-dotted
 curves denote the waveforms derived by the Taylor-T4 formula.}
 \label{fig:waveform}
\end{figure*}

The gravitational waveforms for selected models are plotted in
Fig.~\ref{fig:waveform}. As often found in our series of papers
\cite{kst2010,kost2011}, the waveform agrees well with the Taylor-T4
formula except for the first one orbit and for the final inspiral orbit
just before the merger.  For completeness, corrected results for the
radiated energy and angular momentum are also listed in
Table~\ref{table:gw}.

\begin{table*}[tbp]
 \caption{Corrected numerical results for gravitational waves: Compare
 with Table IV.}
 \begin{tabular}{ccccccccc}
 \hline \hline
 Model & $\Delta E / M_0 (\%)$ & $\Delta J / J_0 (\%)$ & ~(2,2)~ &
 (3,3)~ & ~(4,4)~ & ~(2,1)~ & $f_\mathrm{QNM} M_\mathrm{BH}$ &
 $V_\mathrm{kick}$ (km/s) \\
 \hline
 M20.145 ($N=60$) & 0.80 & 17 & 0.78 & 0.014 & 0.004 & 0.001 & 0.084 &
                                 25 \\
 M20.160 ($N=60$) & 1.13 & 20 & 1.10 & 0.024 & 0.007 & 0.003 & 0.084 &
                                 43 \\
 M20.178 ($N=60$) & 1.62 & 23 & 1.56 & 0.037 & 0.010 & 0.008 & 0.083 &
                                 80 \\
 M30.145 ($N=60$) & 0.99 & 18 & 0.93 & 0.041 & 0.009 & 0.003 & 0.077 &
                                 20 \\
 M30.160 ($N=60$) & 1.37 & 22 & 1.28 & 0.068 & 0.012 & 0.007 & 0.077 &
                                 15 \\
 M30.178 ($N=60$) & 1.70 & 24 & 1.56 & 0.11 & 0.017 & 0.010 & 0.077 &
                                 40 \\
 M40.145 ($N=60$) & 1.09 & 20 & 0.97 & 0.084 & 0.018 & 0.016 & 0.073 &
                                 92 \\
 M50.145 ($N=60$) & 1.01 & 20 & 0.86 & 0.11 & 0.022 & 0.012 & 0.071 &
                                 84 \\
 \hline \hline
 \end{tabular}
 \label{table:gw}
\end{table*}

\begin{figure*}[tbh]
 \begin{tabular}{cc}
  \includegraphics[width=75mm,clip]{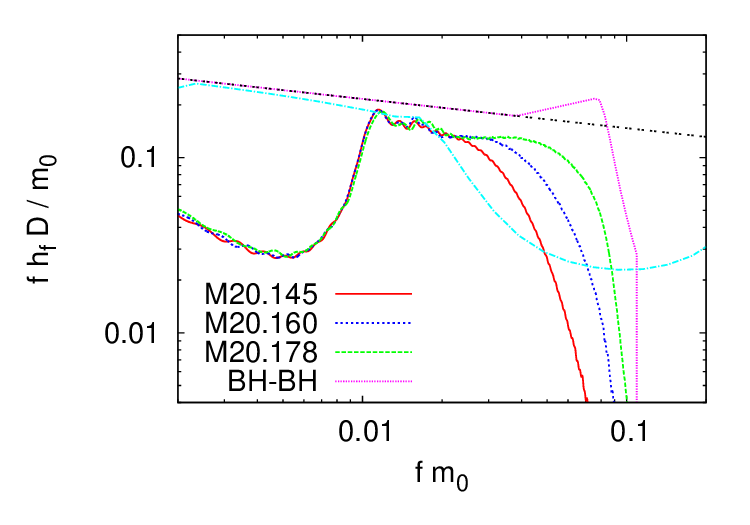} &
  \includegraphics[width=70mm,clip]{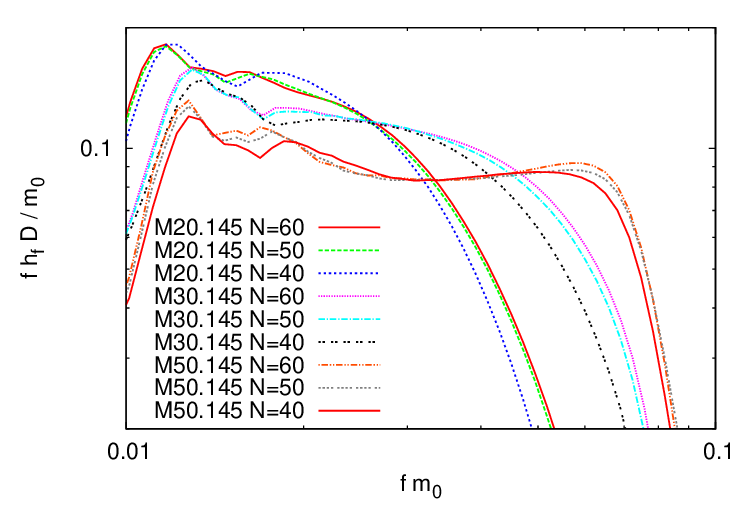}
 \end{tabular}
 \vspace{-4mm}
 \caption{Left: The spectrum of gravitational waves $f h(f) D / m_0$ for
 models M20.145, M20.160, and M20.178 with $N=60$. The dashed and
 long-dashed curves denote spectra computed in the quadrupole and
 Taylor-T4 formulae, respectively. The curve labeled ``BH-BH'' is the
 spectrum of the phenomenological waveform for binary black holes
 proposed in~\cite{abchkswbddghprsst2008}. Right: The spectra for models
 M20.145, M30.145, and M50.145 with different grid resolutions. We do
 not match numerical data to the Taylor-T4 waveform unlike in our
 original version.} \label{fig:spectrum}
\end{figure*}

The gravitational-wave spectra for selected models are shown in the
left panel of Fig.~\ref{fig:spectrum} together with the spectra
computed by the Taylor-T4 formula and of the phenomenological waveform
proposed in~\cite{abchkswbddghprsst2008} following the idea
of~\cite{elsb2009}. We also recomputed the cutoff frequency,
$f_\mathrm{cut}$, for each model, and the results are listed in
Table.~\ref{table:remnant}. We reconfirm the correlation between
$\mathcal{C}$ and $f_\mathrm{cut} m_0$ for the fixed $Q$ (the cutoff
frequency converges approximately for $N=60$ and 50). The right panel
of Fig.~\ref{fig:spectrum} shows the spectrum for selected models with
different grid resolutions, and indicates that the spectrum is in the
convergent regime for $N=60$ and 50.

Finally, we would like to apologize to readers of the original version
for any inconvenience caused by our negligence.

\end{document}